\documentclass[amsmath,amssymb]{revtex4}
\usepackage[dvips]{graphicx}
\usepackage{epsfig}
\usepackage{subfigure}
\newcommand{\ket}[1]{\left| #1 \right\rangle}

\renewcommand{\>}{\rangle}
\newcommand{\be}{\begin{equation}}
\newcommand{\ee}{\end{equation}}
\newcommand{\bea}{\begin{eqnarray}}
\newcommand{\eea}{\end{eqnarray}}

\graphicspath{%
    {converted_graphics/}
    {/}
    {FMOrev/}
}
\begin{document}
\title{Multipartite entanglement  in  the Fenna-Matthews-Olson (FMO) pigment-protein complex}
\author{ A. Thilagam} 
\email{thilaphys@gmail.com}
\address{Information Technology, Engineering and Environment, 
Mawson Institute,
University of South Australia, Australia
 5095} 

\begin{abstract}
We investigate multipartite states in the Fenna-Matthews-Olson (FMO) pigment-protein complex
of the green sulfur bacteria using a Lorentzian spectral density of the phonon reservoir fitted with
typical parameter estimates of the species,  \textit{P. aestuarii}. The evolution of the entanglement measure
of the excitonic $W$ qubit states is evaluated in the picosecond time range, showing increased revivals
 in the non-Markovian regime. Similar trends are observed in the evolution
dynamics of the Meyer-Wallach measure of the $N$-exciton multipartite state, with results showing
that multipartite entanglement can  last from 0.5 to 1 ps, between the Bchls of the FMO complex.
The teleportation and quantum information splitting fidelities associated with the $GHZ$ and $W_A$
resource states of the excitonic qubit channels of the FMO complex show that revivals in fidelities
 increase with the degree of non-Markovian strength of the decoherent environment.
Results indicate that quantum information processing tasks involving teleportation followed by the
decodification process involving $W_A $states of the FMO complex, may play a critical role during
coherent oscillations at physiological temperatures.
\end{abstract}
\maketitle

\section{Introduction}
During photosynthesis,  the vital processes of light absorption, charge separation
 and  efficient energy transfer to a reaction center (RC)  occurs and   culminates
 in the conversion of electronic energy to useful chemical energy \cite{knox,p1,p1a,p3,may,ame}.  
These processes are performed by light harvesting 
(LH) complexes in  photosynthetic systems. The well known purple photosynthetic 
bacteria constitutes two different types of LH system units with ring-like structures,
 the core LH1 complex which surrounds each RC,
and the peripheral LH2  complexes that  transfer energy to the LH1 complex \cite{sun,flem,cog}.
While the LH1 complex has only one absorption band at around 875 nm, 
the LH2 complex may have two absorption bands (800 and 850 nm) in the case of
{\it Rhodobacter (Rb.) sphaeroides}. In the  Fenna-Matthews-Olson (FMO) pigment-protein complex
of the  green sulfur bacteria, sunlight is transported in the form of electronic 
energy by  bacteriochlorophyll (BChl) molecules to the  RC complex \cite{flem2,fenn,cam,engel,reng,wend}.
 Light energy is harnessed with a near  efficiencies of 95\%  \cite{effi} or
 more in natural photosynthetic systems, such efficiencies have yet 
to be realized in artificial light harvesting complexes.  
The  observation of long lived coherences, lasting several picoseconds, in photosynthetic 
antenna complexes \cite{engel,flemB,pani,lee2007,collini,brix05,sav,sens,renrat,van} 
has been linked to quantum entanglement and elements of quantum communication protocols
 during  exciton (correlated electron-hole pair) propagation.  

In general, the exciton propagation times between  the photosynthetic
pigment molecules are about 100 fs.  At room temperatures, biological systems undergo large decoherences
due to interactions with  the surrounding environment, and 
excitonic coherences are not expected to persist at the comparatively longer  times
noted in the experimental results. Partly for this reason, there is theoretical interest in the 
accurate modeling of  the  dynamics of the exciton  to examine the underlying basis for efficient energy transfer  in photosynthetic structures. However, the energy scales involved during the energy delocalization and
decoherence due to lattice vibrations are almost equivalent  
($10-100\ {\rm cm}^{-1}$),  which makes it a challenging task to model
photosynthetic systems. The Markovian approximation in which an infinitely short correlation reservoir
time  is assumed, provides convenience in computational analysis, but yields reliable  results only under certain limiting conditions.  Consequently, the exciton hopping model  \cite{knox,p0,fors} has 
 been replaced by  sophisticated quantum mechanical models of exciton propagation 
 incorporating  quantum coherence features in recent works \cite{lloyd,alex,silbey,ghosh,sar,hierac,shaun}.
 
While the  coupling of the excitonic states to a dissipative 
environment is expected to  lead to fast decoherence, several works have shown that the environment
noise can enhance the propagation of  energy  in 
light harvesting systems ~\cite{lloyd,ok3,plenio08,ishi09,reb09,caru09,caru10}.
Studies \cite{caru09,caru10,strump11,chensil} have shown that attributes such as the  non-Markovian
 interactions and quantum correlations present in the phonon
 reservoir  help improve the  performance 
of the light harvesting complex.
In Ref.\cite{caru09}, Caruso et. al.  showed using  analysis of the  entanglement properties in the  
Fenna-Matthew-Olson complex,   that  the delicate interplay of quantum mechanical features and the 
environmental noise  plays a critical role in bringing about an  optimal system performance.
The authors examined entanglement  in further detail in the FMO complex, under different natural and artificial conditions and Markovian and non-Markovian dynamics \cite{caru10},
and showed that near unit energy transfer occurs at the optimal system operating point
at which the entanglement measure reaches only intermediate values. 
In this regard, a thorough  examination of  the structure and interactions of quantum 
states in the abstract vector space known as the Hilbert space will contribute
 to greater insight of optimal processes in photosynthetic systems. 
 
A significant  feature  of quantum states is  their linear structure in Hilbert space.
Unusual consequences  arise when each quantum state can exist as a sum of two distinctly
different states in the tensor product space of the two separate Hilbert subspaces. 
This unique property underpins many puzzling properties of 
quantum states, including  quantum entanglement which 
 is a valuable resource for the implementation of quantum computation
and quantum communication protocols \cite{niel,horo,horo96}, which  include 
quantum teleportation \cite{ben93,tele2,pop}, 
dense coding \cite{wies}, quantum cryptography \cite{hill99,fuch} and
 remote state preparation \cite{remote}. Moreover, entanglement effects
underpins the emergence of classical macroscopic features in a system undergoing
quantum  mechanical  interactions within itself and with its surrounding environment.
 The degree of bipartite entanglement of  a quantum system  which is composed of two subsystems
is maximum (minimum)  when the constituent subsystems are in  completely mixed states (pure states). The well known concurrence measure \cite{woot} is  zero for
 separable states, and becomes one when maximal ignorance of each qubit  state is achieved in the case of
maximally entangled states. A non classical correlation measure known as
the quantum discord \cite{zu,ve1,ve2}  was shown to be non-zero  even   in non entangled states,
and more recently,  its q-entropies \cite{qentro} was shown to  incorporate aspects of non-classicality
not visible within the conventional  definition of the quantum discord.

The exciton manifest as  delocalized excitations over the 
real crystal space \cite{reng,Davy,Craig,toy,suna,thilaold}  and thus provides
the  ideal example of an extended entangled system in molecular systems \cite{thilapra}.
To this end, the  exciton can be considered to play a central role in the 
entanglement properties of  photosynthetic systems by being 
 in a state of existence at several lattice sites and traversing multiple paths simultaneously. 
The  degree of  exciton delocalization is however limited by the fragility of the 
 coupled  molecular chromophores interacting
with a bath of  environmental phonons and static disorder.
While the underlying feature that an exciton with a wavevector K is delocalized in real space.
is linked with the robustness in quantum coherence during solar energy related biochemical reactions,
the exact details of such investigations has yet to be firmly established
in current research on  exciton propagation in photosynthetic systems .
The biggest obstacle to understanding these  quantum features  is that till now,
 it is not immediately clear as  to how entangled excitonic states  act to maintain quantum coherence in 
photosynthetic processes.

It is important to note the distinction between  entanglement measures and those associated with 
coherences in entangled states which evolve under the influence of
 environmental variables \cite{death1,death2,death3,dur,aol}. 
A global system constituting several subsystem possess  entanglement which decays
in ways which are vastly different from the coherence of each subsystem, 
when immersed in a decoherent  environment.
This stems from the fact that the global characteristics of entangled systems and their evolution under
local decoherence differ from those of its constituent subsystems.
 Entanglement may vanish at a certain  finite time  prior to the decay of the coherences
\cite{death1,death2,death3,dur,zhong}, and is a topic of interest as the use of 
entanglement as a resource is meaningless if it vanishes rapidly.
 The abrupt disappearance of entanglement at a finite time has been termed entanglement sudden 
death  by Yu et. al. \cite{death1}. In general, this  finite time is 
expected to decrease for multipartite states or $N$-qubit states ($N > 2$) where several   particles are entangled. 
The relationship between coherence and entanglement remains subtle as decoherence  in open quantum
systems takes an infinite amount of time to vanish unlike entanglement, however the degree of robustness of the initial state of the quantum system under study 
is a  critical factor that could link entanglement and 
coherence properties. For instance, it is highly likely that
robustly entangled states as initial states are more resistent to decoherence processes. 

Unlike bipartite states,  investigations  related to the robustness of quantum states 
are    challenging for multipartite states,
where there is still lack of consensus on the categorization and specification of entanglement itself. 
Recent works \cite{gohn,borras} have shown  variations in the robustness of different types 
of multipartite states (Dicke, GHZ, W and cluster states) under decoherence processes.
To this end, there has been scant investigations of the 
characteristics of multipartite states in photosynthetic systems, specifically with respect
to the influence of  the environmental bath on  multipartite excitonic states.
The study of multipartite states  is particularly relevant 
for light harvesting systems, as these states
 possess a  richer source of local and nonlocal correlations due to the
multitude of partitions available within a group of entangled qubits. Moreover there
is  relevance  in the realistic situation  of an entire photosynthetic membrane constituting 
many FMO complexes and thousands of bacteriochorophylls, possibly giving rise
to  large cluster of massively entangled excitonic qubits.
In this regard, tt would be worthwhile to  seek  a detailed understanding
of  how quantum communication protocols \cite{niel,horo} such as  quantum teleportation \cite{ben93,tele2}, 
are utilized by multipartite states   to assist energy transfer in 
photosynthetic processes.  The current work is aimed at addressing some of these issues.

In this work, we consider 
multipartite entanglement in photosynthetic systems and the decay of these states
under decoherence  associated with a phonon bath.
A characterization of the different types
of  multipartite  entanglement  will be examined, including a brief discussion 
of the robustness of the different types of entanglement against external noise 
in open quantum system. The influence of specific tasks such as
 teleportation \cite{ben93,tele2,hotta} and   quantum state splitting \cite{split1,split2}
in a noisy photosynthetic system will be examined as well. 
This work is organized as follows.  In Sec. \ref{mult}, we discuss 
various types of multipartite states and the entanglement quantifiers that are
used to characterize some well known multipartite states. In particular, we discuss in detail
the global multipartite state entanglement measure, which will be subsequently used to evaluate 
entanglement quantifiers specific to  the  FMO complex of  \textit{P. aestuarii}
in Sec. \ref{fmo}. A description of the photophysics and reaction mechanisms
in the Fenna-Matthews-Olson (FMO) complex from the green sulfur bacteria
of the two species, \textit{Prosthecochloris (P.) aestuarii}  and \textit{Chlorobium (C.) tepidum} 
is provided in Sec. \ref{fmo}. In this Section, we also include a detailed evaluation of the exciton-phonon
dynamics based on the time-convolutionless (TCL) projection operator 
technique and a Lorentzian spectral density of the  phonon  reservoir associated with \textit{P. aestuarii}.
Numerical results of the  entanglement measure of  the  excitonic $W$ states of the FMO complex species
 \textit{P. aestuarii} are presented in 
Sec. \ref{wfmo} for both the Markovian and non-Markovian regimes.
A similar analysis is performed with respect to the 
Meyer-Wallach  measure of excitonic qubits in the FMO complex in Sec. \ref{MWfmo}.
In Sec. (\ref{tele}), the teleportation and quantum information splitting 
fidelities of excitonic qubits associated with 
 $GHZ$ and $W_{A}$ resource states, are calculated  for the  FMO complex species
 \textit{P. aestuarii} and the conclusion is  provided in  Sec. (\ref{con}).

\section{Multipartite states and  entanglement measures}\label{mult}

 A  qubit state $\Psi$ associated with $N > 2$ subsystems can be written in the 
multi-qubit state or multipartite state as 
\be
|\Psi\> = \sum_{n=1}^N c_{n} \left( |0\>^{\otimes (n-1)}\otimes|1\>\otimes|0\>^{\otimes (N-n)} \right)
\ee
 with  coefficients  $c_n$.  
Multipartite systems   generally involve  the class of  permutation-symmetric states which are invariant when  any pair of 
particles are swapped, and  include
the Greenberger-Horne-Zeilinger (GHZ) states \cite{green,carb,kempe} and Dicke states \cite{dicke}.
For a system of $N$ qubits, the Dicke states are obtained using the  permutations of  basis states with
$N-K$ qubits being in the $|0\rangle$ state and $K$ in the  $|1\rangle$ state using the notation $| S_{N,K} \rangle$
\be
\label{dicke}
  | S_{N,K} \rangle = {\binom{N}{K}}^{- 1/2} \sum_{\text{perm}} \;
  \underbrace{ | 0 \rangle \otimes | 0 \rangle \otimes \cdots | 0 \rangle }_{N-K}
  \underbrace{ | 1 \rangle \otimes | 1 \rangle \cdots | 1 \rangle }_{K}
  \enspace ,
\ee
where $0 \leq K \leq N$. A well known example of a symmetric
 Dicke state with just one excitation is  the N-qubit $W$ state \cite{cirac}
\be 
\label{wn}
|W_N \> =\frac{1}{\sqrt{N}} \left(|100...0\>+|01...0\>+...+|0...01\> \right),
\ee
The $W$ states are inequivalent to the $GHZ$ states \cite{dur,carb,kempe} which are obtained as
\bea
\left\vert GHZ_{N}\right\rangle &=&\alpha \underbrace{ | 0 \rangle \otimes | 0 \rangle \otimes \cdots | 0 \rangle }_{N}
+\beta \underbrace{ | 1 \rangle \otimes | 1 \rangle \otimes \cdots | 1 \rangle }_{N} \\ \nonumber
&=&
\alpha\left\vert 0\right\rangle ^{\otimes N}+\beta\left\vert 1\right\rangle ^{\otimes N}
\eea
where  $\left\vert \alpha\right\vert
^{2}+\left\vert \beta\right\vert ^{2}=1$. 
States  derived from each other via local operations and
classical communication (LOCC) belong to the same group 
of equivalent resources for  quantum information tasks such as quantum
teleportation \cite{niel}. Consequently, there are  differences in robustness against  decoherence
between entangled $W$  and $GHZ$ states. The $W$ state  is known to be 
highly robust in its genuine multipartite  entanglement with respect to loss of a single excitation \cite{cirac}.
However  it is still not conclusive whether the  $W$ state is more  robust against decoherence
than the $GHZ$ states, mainly due to the incompatibility in comparison of the dynamics of the two types
of states during decoherence. It has been shown that in the case of
 $GHZ$ states undergoing decoherence \cite{aol}, the 
entanglement decay occurs faster when the number of initially entangled particles
increases.

Bipartite entanglement are generally quantified by measures such as 
the von Neumann entropy \cite{kitaev,batle}, negativity \cite{horo96,nega}, concurrence \cite{woot} and
quantum discord \cite{zu,ve1,ve2}. 
These measures cannot be extended to the case of
multipartite system without introducing  an element of subjectivity associated with the partitioning
of the global system  into smaller systems. As a consequence  different entanglement  measures 
 are expected to  display  variations in their decay under decoherence. 
Currently,  there exists several entanglement measures, such as the 
the Meyer and Wallach  measure \cite{meyer} and its extension based on the normalized negativity 
 which are  used to quantify multipartite entanglement. 
In a recent work,  Chaves et. al. \cite{chaves} 
 identified entanglement measures of states undergoing
decoherence with tasks such as  teleportation
\cite{ben93,tele2} and the splitting \cite{split1,split2} of quantum
information.  Thus in this elegant approach, a task associated with 
quantum information processing was associated with the entanglement 
quantifiers.

A commonly adopted approach that is used to determine the multipartite entanglement measure
of  a $N$-qubit state first  involves  partitioning it 
$2^{N-1}-1$  times. Subsequently, a bipartite  entanglement measure
such as the von Neumann entropy \cite{kitaev,batle} or negativity \cite{horo96}  can be 
evaluated for  each bi-partitioned density matrix.
The average of the sum of the entropies or negativities over all possible partitions
may be used to characterize the global entanglement. 
In connection with this, we adopt
 the global multipartite state entanglement measure used in Ref.\cite{borras}
\bea
\label{ent}
E &=& \frac{1}{[N/2]} \sum_{m=1}^{[N/2]} E^{(m)}, \\
E^{(m)} &=& \frac{1}{N_{bi}^m} \sum_{i=1}^{N_{bi}^m} E(i).
\label{Entsub}
\eea
where $E(i)$ denotes the entanglement associated with
a single bipartition of the $N$-qubit system and 
$E^{(m)}$ denotes the average entanglement between subsets of $m$
qubits and the remaining $N-m$ qubits. The
average is performed over the $N_{bi}^{(m)}$ nonequivalent
bipartitions, $ N_{bi}^{m} = \binom{N}{n}\textrm{     if }n \neq N/2$ and
$N_{bi}^{N/2} =\frac{1}{2} \binom{N}{N/2}\textrm{     if }n = N/2$.
The total number of bipartitions is equal to $N_{cuts} = \sum_{i=1}^{[N/2]} N_{bi}^{i} = 2^{N-1}-1$.
Due to the occurrence of  mixed states,  the negativity is employed as a suitable measure 
of bipartite entanglement. The normalized negativity is defined as
\cite{maj}
\begin{equation}
\label{nnega}
E(i)= \frac{2}{2^m-1}\sum_i|\alpha_i|,
\end{equation}
where $\alpha_i$ is the negative eigenvalue of the partial
transpose matrix for $m$ and the remaining $N-m$ bipartition.
The measures in  Eqs. (\ref{Entsub}) and (\ref{nnega}) will be employed in the next section
to evaluate the entanglement dynamics of the excitonic qubits in the FMO complex.

\section{Excitonic qubits in the FMO complex}\label{fmo}
\begin{figure}[htp]
  \begin{center}
\subfigure{\label{aa}\includegraphics[width=8cm]{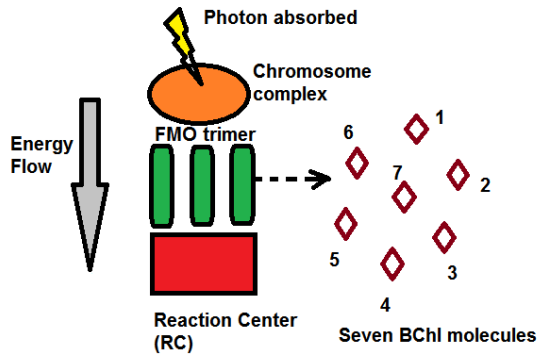}}\vspace{-1.1mm} \hspace{1.1mm}
     \end{center}
  \caption{Simplified scheme of  the  light reaction  in which 
photons are absorbed by a network of  antenna pigments linked
to the FMO complex, and  from which  the excitation energy is transported
to the reaction center (RC). Photochemical reactions take place at the RC
 to convert the excitation energy into chemical energy.}
\label{scheme}
\end{figure}

In this Section, we examine the exciton entanglement dynamics of the 
 Fenna-Matthews-Olson (FMO) complex from the green sulfur bacteria
of the two species, \textit{Prosthecochloris (P.) aestuarii}  and \textit{Chlorobium (C.) tepidum} 
based on semiempirical values of the reservoir characteristics at 
cryogenic temperatures  \cite{lorenExpt}. 
Fig.~\ref{scheme} shows a simplified  light reaction scheme in which 
photons are absorbed by a network of  antenna pigments linked
to the FMO complex, and  from which  the excitation energy is transported
to the reaction center (RC) where photochemical reactions 
 convert the excitation energy into chemical energy.
The FMO complex  trimer, which constitutes three identical monomer subunits
with  each unit constituting seven bacteriochlorophyll (BChl)a molecules, 
 acts as an efficient channel of energy propagation from the light harvesting molecular 
complex to the  reaction center.

In the experimental work \cite{lorenExpt}, FMO complexes from 
 \textit{Prosthecochloris (P.) aestuarii}  and \textit{Chlorobium (C.) tepidum} were isolated 
and dissolved in suitable solvents before time-resolved transient 
absorption measurements were performed. The site energies and coupling 
between  each  BChl\textit{a} molecules of  \textit{Prosthecochloris (P.) aestuarii}  
and \textit{Chlorobium (C.) tepidum} were characterized
Ref. \cite{lorenExpt,louwe,vulto2} using an effective exciton model  \cite{louwe},
which accounts well for the optical steady state spectra and the dynamics of the  FMO complexes.
For both species of FMO complexes, the third BChl (numbered 3 in Fig.~\ref{scheme})
was found to have  the lowest site energy  \cite{lorenExpt}. However there are subtle differences
between the two species as   the absorption
spectrum of \textit{P. aestuarii} has an absorption spectrum maximum height at 815 nm,
while that of the \textit{C. tepidum} has its maximum at 806 nm.
In \textit{P. aestuarii}, BChls 1 and 4 contribute to the 
815 nm  absorption spectrum band whereas the contribution
to the same band arises mainly from BChl 4 in \textit{C. tepidum}.
Consequently, we expect differences in the coherence features that  can
arise when these two species are coupled to a phonon bath with 
similar environmental parameters.

\begin{table}
\begin{center}
\caption{
\label{tbl}
The Bchl site energies (cm$^{-1}$) used in 
Refs. \cite{reng} (provided as true values of the site energies
in last column of Table 3),  \cite{wend} and \cite{lorenExpt} for the 
FMO complex of  \textit{P. aestuarii}. The difference in site energies (provided within \{ \}) is calculated
by considering the site energy of the third BChl as a reference point. 
}
\begin{tabular}{@{}llllllll}
\\
\hline \hline
 BChl \quad \quad \quad&&Exciton energy (cm$^{-1}$) \cite{lorenExpt} &&Exciton energy (cm$^{-1}$)  \cite{reng}   &&Exciton energy (cm$^{-1}$)  \cite{wend} \\ 
\hline \hline
3&& 12,112 \{0 \}&& 12,210  \{0\} && 12,175  \{0\} \\
1 &&12,266  \{154\} &&12,450 \{240\} && 12,315  \{140\}  \\
4 && 12,293 \{181\} &&12,320 \{110\} && 12,405  \{230\} \\
6&&12,396 \{284\} && 12,540 \{330\} && 12,430  \{255\} \\
7&& 12,457 \{345\} &&12,470 \{260\} && 12,450  \{275\}\\
2&&12,496 \{384\} && 12,520 \{310\} && 12,500 \{325\}\\
5&&12,634 \{522\} && 12,550 \{340\} && 12,625  \{450\}\\
\hline
\hline
\end{tabular}
\end{center}
\end{table}

 In Table  \ref{tbl}, the Bchl site energies for the FMO complex of  \textit{P. aestuarii}
(cm$^{-1}$) given in  Refs. \cite{reng},  \cite{lorenExpt} and  \cite{wend} are provided.
The difference in site energies (provided within \{ \}) is calculated
via subtraction  from the site energy of the third BChl. 
It is important to distinguish the  propagating exciton 
 from the presence of excitation at a specific Bchl site, as it is well
known that an exciton with a wavevector ${\bf K}$ is delocalized in real space.
The  qubit state has a probabilistic occupation at all  the  BChl
sites, with one or two BChl sites contributing predominantly to the qubit state.
 We use  an effective exciton Hamiltonian defined in the basis of the Q$_y$
one-exciton states in which  the   Q$_y$ bandwidth is associated with 
the lowest excited state of the BChl molecule.  
When  a photon is absorbed by the BChl molecule,
 the newly created exciton propagates rapidly to the adjacent units while interacting 
with a bath of phonons. The exciton can be considered as delocalized over all the seven BChls
so that the  excitonic  states are obtained as 
 linear combinations of the excited state wave functions of
the individual  BChl.  In this regard, a  simple way  to represent excitonic states as qubit states 
is by  associating a qubit with  the presence (or absence) of an exciton at a
specific energy level. Each energy level can be modeled as a two-level system,
with $\ket{0}_{e_{_1}}$ ($\ket{1}_{e_{_1}}$) denoting 
the absence (presence) of an excitation at the specified energy level.

In order to compute the  contributions
from different pigments to  the seven exciton states,
we  model the subunit of the FMO complex 
via the  Hamiltonian, $\hat H_{ex}$  in the site basis,
with the coupling energy terms given  in  units of cm$^{-1}$
\begin{equation}\label{FMOcom} \hat H_{ex}= \left(\begin{array}{ccccccc}
240.0& -104.1& 5.1& -4.3 & 4.7& -15.1& -7.8\\
-104.1& 310.0& 32.6& 7.1& 5.4& 8.3& 0.8\\
 5.1& 32.6& 0& -46.8& 1.0& -8.1& 5.1\\
 -4.3& 7.1& -46.8& 110.0& -70.7& -14.7& -61.5\\
 4.7& 5.4& 1.0& -70.7& 340.0& 89.7& -2.5\\
 -15.1& 8.3& -8.1& -14.7& 89.7& 330.0& 32.7\\
 -7.8& 0.8& 5.1& -61.5& -2.5& 32.7& 260.0\\
\end{array} \right). \end{equation}
The off-diagonal terms in $\hat H_{ex}$ are obtained  using
 the  intersite excitonic couplings  between various BChls as
given  in Ref.\cite{reng} for the FMO complex of  \textit{P. aestuarii}. 
The various coupling energies were evaluated  \cite{reng}
 in a dielectric environment by solving  a Poisson equation for each BChl  by a finite difference method.
The diagonal terms of $\hat H_{ex}$ in Eq. (\ref{FMOcom})
were obtained using the energy differences for the respective BChl site using the
site energies, provided within \{ \},  in Table  \ref{tbl}.
The exciton energies and   eigenstates of $\hat H_{ex}$  in Eq. (\ref{FMOcom}) can be evaluated  using
standard diagonalization techniques. The seven excitonic qubits 
(labelled in order of increasing energy) are given in Table  \ref{tblqu}.
It can be seen that each exciton state is generally associated with 1-3 Bchl sites,
and  the lowest energy exciton qubit state $\ket{1}_{e_{_1}}$ with energy $-24$ cm$^{-1}$  is localized mostly on BChls 3 and 4. 
Most importantly, Table  \ref{tblqu} shows that each exciton qubit state may have simultaneous occupation of
two or more BChl sites, a feature that is only feasible in  quantum systems.
\begin{table}
\begin{center}
\caption{
\label{tblqu}
Delocalized exciton qubit states provided
as linear combinations of   probability occupation amplitudes associated with
the seven BChl sites. The excitonic qubit states are labeled according to increasing  exciton energies for the
FMO complex of  \textit{P. aestuarii}.  
}
    \begin{tabular}{| r | r | r | r | r  | r |  r |  r | r |}
    \hline
     & ~ $\ket{1}_{e_{_1}}$& ~ $\ket{1}_{e_{_2}}$& ~ $\ket{1}_{e_{_3}}$ & ~ $\ket{1}_{e_{_4}}$ & ~ $\ket{1}_{e_{_5}}$ & ~ $\ket{1}_{e_{_6}}$& ~ $\ket{1}_{e_{_7}}$   \\ \hline 
$E_x$ (cm$^{-1}$) & -24 & 86  &167 & 251 & 280 & 385 & 444   \\ 
  BChl 1 \quad \quad& 0.0553 & 0.0750 & 0.8081 & -0.0444 & -0.0305 & 0.5688 & 0.1087 \\
  BChl 2 \quad \quad & 0.1155 & 0.0609 &0.5555 & -0.1127 & -0.0950 &-0.7943 & -0.1475 \\ 
  BChl 3 \quad \quad & -0.9062 & -0.3811 & 0.1314 &-0.1119 & 0.0214 & -0.0515 & -0.0232  \\ 
  BChl 4 \quad \quad & -0.3903& 0.8121 &-0.0124 & 0.3431   & -0.1501 & -0.0722 & 0.2061  \\ 
  BChl 5 \quad \quad & -0.0730 & 0.2644 & -0.0992 &-0.4529 & -0.4445 & 0.1860 & -0.6911 \\ 
  BChl 6 \quad \quad & -0.01267 & -0.1020 &0.1048 & 0.7200 & 0.2077 & 0.0561 &  -0.6432 \\ 
  BChl 7 \quad \quad &  -0.0662& 0.3249 & 0.0081 & -0.3627 & 0.8522 & 0.0038 &  -0.1793  \\ \hline
    \end{tabular}
\end{center}
 \end{table}

In general, the dimer model consisting of pair of spin-boson  \cite{thilchem} can be  
used to simplify analysis of the entanglement dynamics of the excitonic qubits.
The spin-boson system is a well known quantum dissipative system which 
constitutes  two energy levels that couple   to an infinite system of harmonic oscillators.  The case of the  
multipartite entanglement can be examined by extending the global system to several
spin chains. We model each  two-level excitonic qubit  interacting with
phonons  via the Hamiltonian,
\begin{equation}
\label{ham}
{H}=\omega_0{\sigma}_+{\sigma}_- \sum_{\bf q} \hbar \omega_{\bf q} \,
b_{\bf q}^{\dagger}\,b_{\bf q} + \sum_{\bf q}   \lambda_{_{\bf q}} \,
 \left ( {\sigma}_- \,b_{\bf q}^\dagger +{\sigma}_+ \,b_{\bf q} \right ),
\end{equation}
where ${\sigma}_+=|1\rangle\langle0|$ and
${\sigma}_-=|0\rangle\langle1|$, 
are the respective Pauli raising and
lowering operators of the exciton with   transition frequency, $\omega_0$.

In order to utilize  Eqs. (\ref{Entsub}) and (\ref{nnega}) for further analysis
of the excitonic qubits, and to  keep the problem tractable,
we assume that each qubit is coupled to its own reservoir of phonons.
The phonon reservoir is given by the second term on the right hand side of  Eq. (\ref{ham})
where  $b_{\bf q}^{\dagger}\,$ and $b_{\bf q}\,$
are the respective  phonon creation and  annihilation 
operators with wave vector ${\bf q}$. The last term of Eq. (\ref{ham})
denotes  the qubit-oscillator interaction Hamiltonian which we assume to 
be linear in terms of the phonon operators. 
 $\lambda_{_{\bf q}}$ is the coupling between the qubit and the
environment and is characterized by the spectral density function,
$J(\omega)=\sum_{\bf q}\lambda_{_{\bf q}}^2\delta(\omega-\omega_{\bf q})$.

For  the spectral density associated with the  coupling of the
 BChl exciton to the phonon bath  environment, we consider
 a Lorentzian spectral density of the reservoir fitted with estimates
derived from site-selective fluorescence measurements on the 825
nm band of the FMO complex of  \textit{P. aestuarii}  \cite{lorenExpt}
  \be 
\label{spectral}
J(\omega)=  \frac{1}{2\pi }\gamma_0 \; \left( \frac{\Delta \omega}{2} \right)^2
 \frac{1}{(\omega_0  - \delta -\omega)^2+\left( \frac{\Delta \omega}{2} \right)^2}
\ee
whose  central peak is detuned from the exciton transition frequency
$\omega_0$  by an amount $\delta$ which is treated as a parameter for subsequent
analysis.  It important to distinguish between the molecular transition frequency 
associated with the $Q_y$ band of the Bchl-a molecular system (about 12100cm$^{-1}$, see
 Table  \ref{tbl} )
and the actual excitonic energies which are of the order of the typical binding energies of
0.1-0.2 eV in organic molecules \cite{Craig}. 
We  consider that the phonon modes  have maximum energies
 just enough to convert the exciton into free electron-hole pairs,
however in general the  greater involvement of a lower range of
 phonon energies of the order 0.01 eV (or 80 cm$^{-1}$)
is assumed. 
As a standard, we use  the full width at half-maximum 
$\Delta \omega$=80 cm$^{-1}$ as used in Ref.\cite{lorenExpt},  noting that the width
is related to the  reservoir correlation time $\tau_B$ via 
$\Delta \omega = \frac{2}{ \tau_B}$. However in 
 order to examine the influence of the  reservoir correlation time $\tau_B$
on population differences, we parameterize $\Delta \omega$ around the region
where phonon energies lie close to 80 cm$^{-1}$.

The parameter $\gamma_0$ is associated with 
 the relaxation time scale $\tau_R$ of the exciton via  the relation
$\tau_R=\gamma_0^{-1}$, and  $\gamma_0$=111ps$^{-1}$  was used in Ref. \cite{lorenExpt}.
 $\gamma_0$, is expected to increase with temperature,
hence for subsequent calculations, this term will be treated as a parameter.
While  the conditions under which the experimental work
in  Ref. \cite{lorenExpt} was carried out precludes
significant temperature-dependent
 dephasing processes, we consider higher values of $\gamma_0$  
(1000 to 2000 cm$^{-1}$) to incorporate non-Markovian effects. 
Hence we assume that the spectral density profile in  Eq. (\ref{ham})
holds valid at higher  temperatures ($>50$ K).

 Due to the exciton-phonon interactions,  the excitonic  qubit decays  to 
oscillator states in the reservoir, making  a transition from  its excited state
$\ket{1}_{e}$ to ground state $\ket{0}_e$.  We 
assume an initial state of the qubit with  its corresponding 
reservoir in the vacuum state of the form \cite{thilazeno}
\begin{equation}
|\phi _{i}\rangle =|1\rangle _e\otimes
\prod_{k=1}^{N'}|0_{k}\rangle _{\mathrm{r}}=
|1\rangle _e\otimes
\ket{{ 0}}_{\mathrm{r}}, 
 \label{initial}
\end{equation}
where $\ket{{ 0}}_{\mathrm{r}}$ implies that all $N'$ wavevector modes 
of the reservoir are unoccupied in the initial state.
The subscripts $e$ and $r$ refer to the excitonic qubit and the corresponding reservoir
respectively.  $|\phi _{i}\rangle$    undergoes subsequent decay of the 
following form
\begin{equation}
|\phi _{i}\rangle \longrightarrow
 u(t) \; \ket{1}_e
\ket{{ 0}}_{\mathrm{r}} + v(t) \; \ket{0}_e
\ket{{ 1}}_{\mathrm{r}} ,  
\label{fstate}
\end{equation}
In order to keep the problem tractable we
consider that $\ket{{ 1}}_{\mathrm{r}}$ denotes
a  collective state of  the reservoir as follows
\begin{equation}
\label{crstate}
|{ 1}\rangle _{\mathrm{r}}=\frac{1}{v(t)}
\sum_{n} \lambda _{\{n\}}(t)|\{n\}\rangle ,
\end{equation}
where $\{n\}$ denotes an occupation scheme in which there are
$n_i$ oscillators with wavevector $k=i$ in the reservoir
and we define the state $|\{n\}\rangle$ as \cite{thilazeno}
\begin{equation}
\label{eq:scheme}
|\{n\}\rangle =|n_0,n_1,n_2...n_i..n_{N'}\rangle ,
\end{equation}
In the  collective reservoir state,  the phonon oscillators
can be  present at all allowed modes, including
simultaneous excitation of several  phonon states. 
At non-zero temperatures, the reservoir state is a Boltzmann 
weighted average over all possible permutations of the
occupation scheme $\{n\}$. 

Following the time-convolutionless (TCL) projection operator 
technique described in Chapter 10 of  Ref.\cite{breu}, we obtain
an expression for the coefficient $u(t)$ (see  Eq. (\ref{initial})),
\begin{equation}
\label{laplace}
\dot{u}(t)=-\int\limits_0^t \; \int {d\omega J(\omega )} \exp
[i(\omega_0-\omega)(t - t_1 )] \; u(t_1)dt_1,
\end{equation}
which can be solved by means of a Laplace transformation. 
The time-convolutionless  projection operator 
technique is independent of the choice of the  spectral density of the environment 
and accounts for strong non-Markovian dynamics \cite{breu},
that may arise in  the photosynthetic molecular environment.

\begin{figure}[htp]
  \begin{center}
\subfigure{\label{aa}\includegraphics[width=5.5cm]{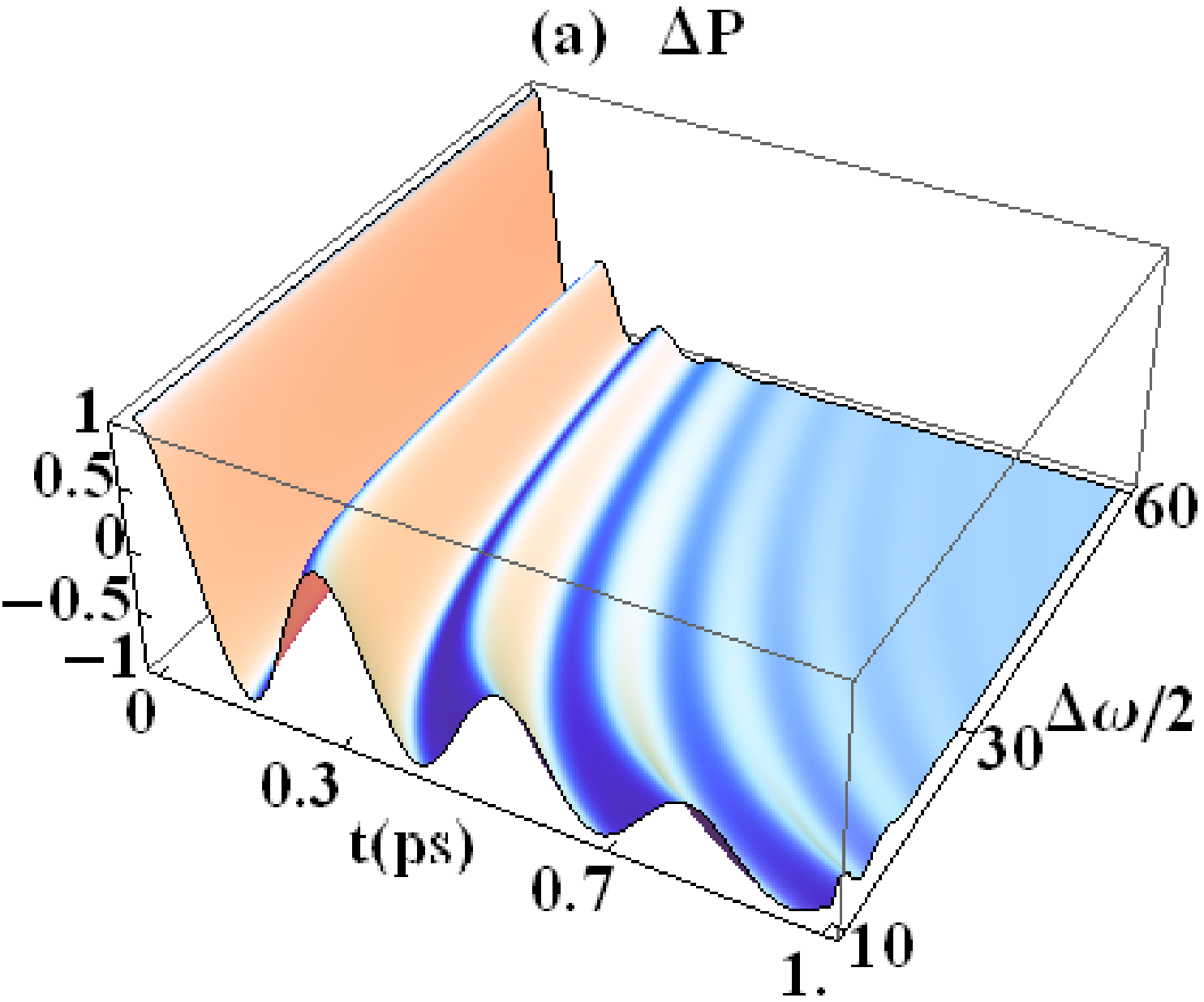}}\vspace{-1.1mm} \hspace{1.1mm}
\subfigure{\label{bb}\includegraphics[width=5.5cm]{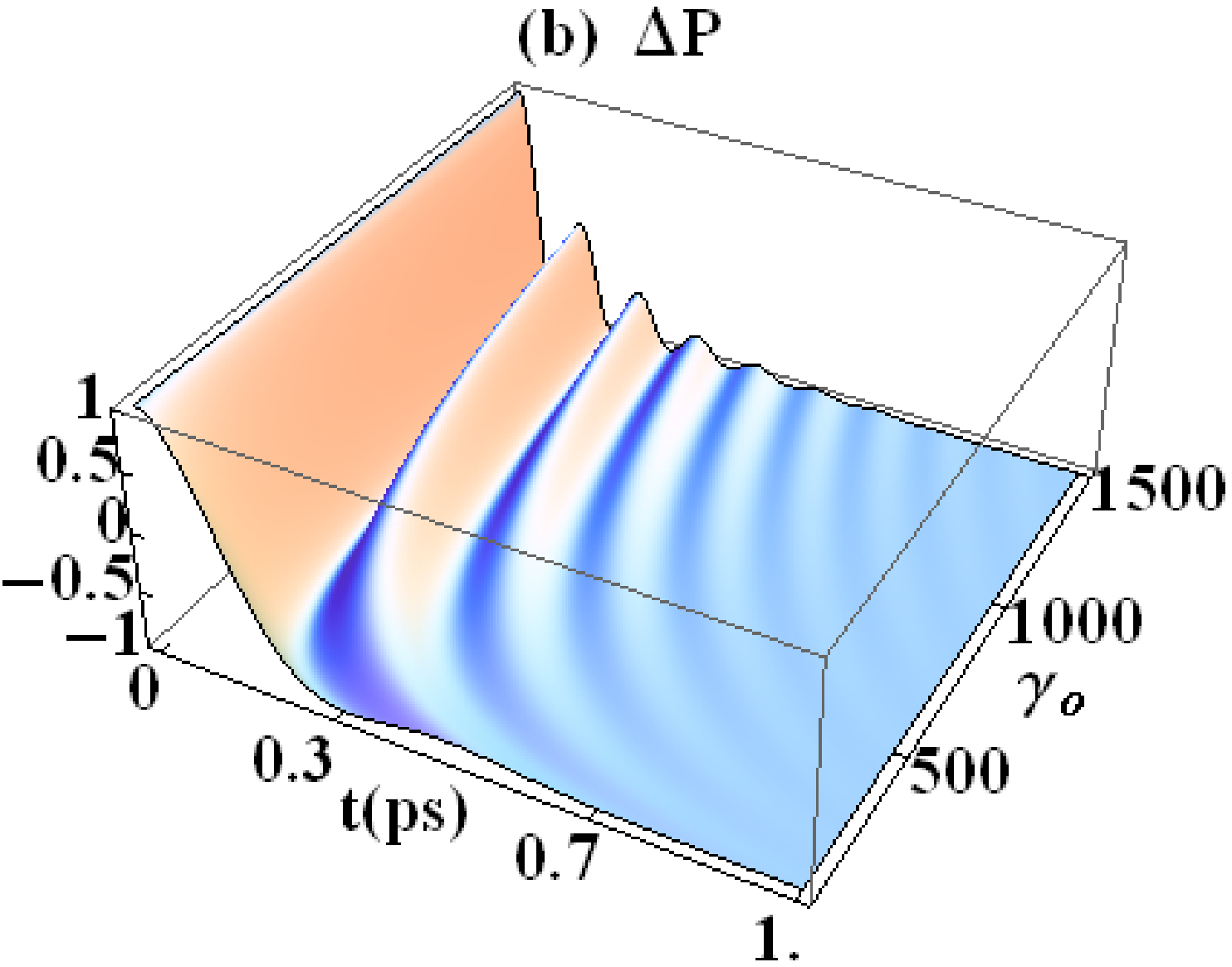}}\vspace{-1.1mm} \hspace{1.1mm}
\subfigure{\label{bb}\includegraphics[width=5.5cm]{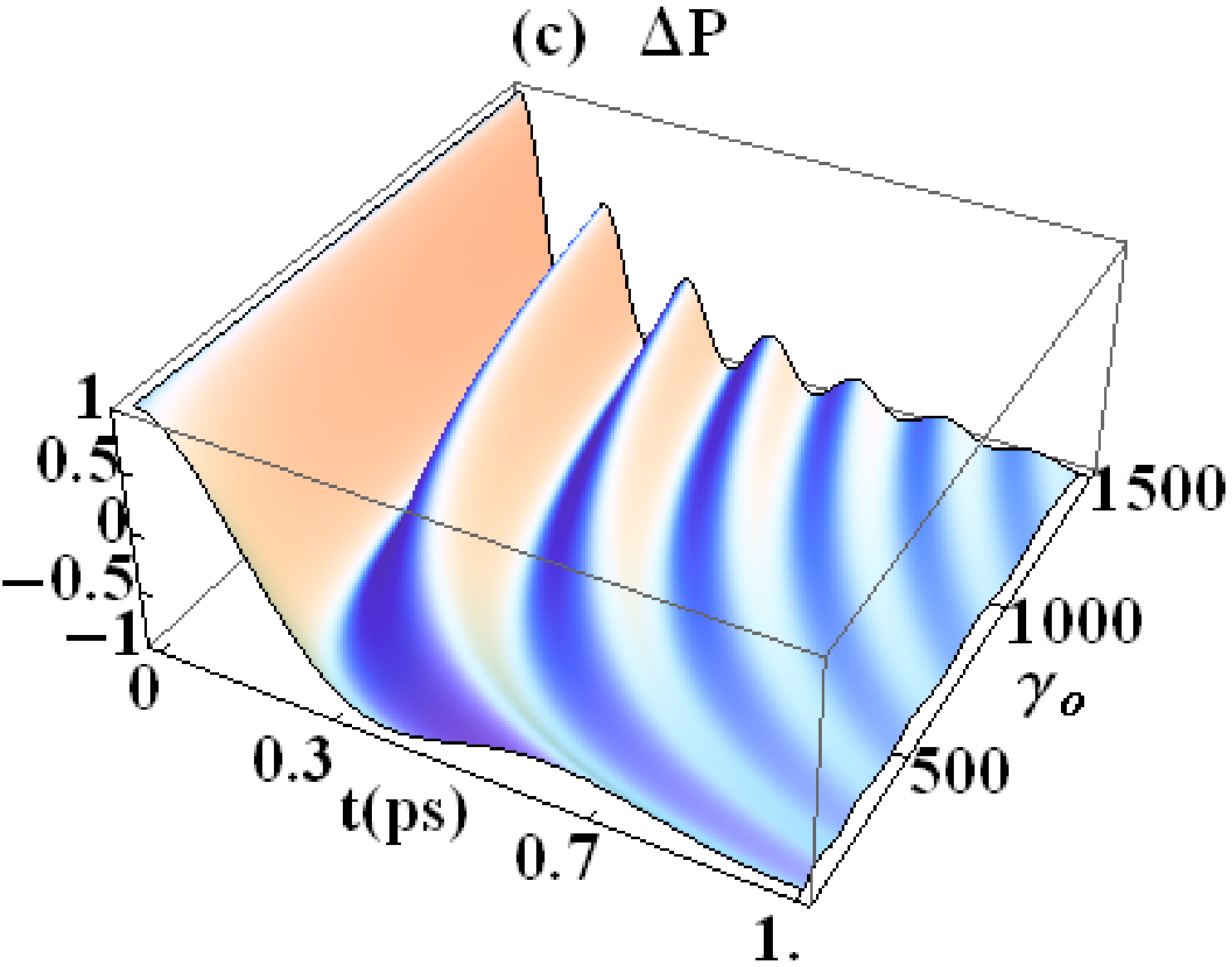}}\vspace{-1.1mm} \hspace{1.1mm}
     \end{center}
  \caption{(a) The  population difference, $\Delta P$ between the 
excited state $\ket{1}_{e}$ and ground state $\ket{0}_e$, evaluated using $u(t)^2$ (Eq. (\ref{ana}),
as a function of time $t$ (ps)  and $\Delta \omega/2$ (cm$^{-1}$) at  $\gamma_0$ =1000 cm$^{-1}$
and  detuning parameter, $\delta$ =0. \quad (b)  $\Delta P$ as a function of time $t$ (ps)  and $\gamma_0$ (cm$^{-1}$) at $\Delta \omega/2$ =40 cm$^{-1}$. \quad (c) $\Delta P$ as a function of time $t$ (ps)  and $\gamma_0$ (cm$^{-1}$) at $\Delta \omega/2$ =20 cm$^{-1}$. 
}
\label{pop}
\end{figure}

 Eq. (\ref{laplace}) is further simplified 
using the form of spectral density in Eq. (\ref{spectral})
\begin{equation}
\label{laplace2}
\dot{u}(t)=-\int\limits_0^t \; \frac{1}{2} \gamma_0 (\Delta \omega/2- i\delta)
e^{-(\Delta \omega/2- i\delta)|t-t_1|}
\; u(t_1)dt_1,
\end{equation}
The  Laplace transformation of Eq. (\ref{laplace})
 with initial condition $u(0)=1$, yields the solution \cite{breu}
  \begin{equation}
\label{ana}
u(t) = e^{-(\Delta \omega/2- i\delta)t/2)}\left [ \cosh \left (\frac{\xi t}{2} \right) 
+ \frac{\Delta \omega/ 2 - i\delta }{\xi }\sinh \left ( \frac{\xi t}{2} \right )\right],
\end{equation}
where $\xi=\sqrt{(\Delta \omega/2-i\delta)^2-\gamma_0 \;\Delta \omega}$.
Fig.~\ref{pop}a,b,c illustrates the  evolution of
 the  coherence properties as reflected in the 
exciton population difference, $\Delta P$ between the 
excited state $\ket{1}_{e}$ and ground state $\ket{0}_e$, evaluated using $|u(t)|^2$ (Eq. (\ref{ana}).
 The change of $\Delta P$  with time $t$ (ps),   $\Delta \omega/2$  and  $\gamma_0$
shows that for smaller  $\Delta \omega/2 \sim$20 cm$^{-1}$ (or larger
reservoir correlation time) and large $\gamma_0$ (or small exciton relaxation
times), there is increased time period (up to 1 ps)  
 over which the population difference, $\Delta P$ which is a signature 
of coherence, is maintained for a select range of parameters.
The time duration of  effective coherence (about 1 ps)  obtained here
 appear to compare well with times of population transfer
between various exciton levels (about 1 ps for levels 1 and 2) noted 
for the  FMO complex of  \textit{P. aestuarii}  \cite{lorenExpt}.
These results suggest that the non-Markovian
regime is best suited for occurrence of  large coherence times.
The analytical form for $u(t)$ in Eq. (\ref{ana}) 
will be useful for  further analysis of the entanglement
measure for the $W$ states of the FMO complex which  we  describe next.

\subsection{Entanglement dynamics of excitonic $W$ states of the FMO complex}\label{wfmo}
We consider a system of four  non-interacting excitonic qubits
coupled to uncorrelated reservoirs and which are  initially prepared in  the $W$ state as follows
\bea
\label{wini} 
|\phi_0\rangle &=&
(|0001\rangle_{e_{_1}e_{_2}e_{_3}e_{_4}}+|0010\rangle{e_{_1}e_{_2}e_{_3}e_{_4}}
+|0100\rangle_{e_{_1}e_{_2}e_{_3}e_{_4}} \\ \nonumber
  ~&~& +|1000\rangle_{e_{_1}e_{_2}e_{_3}e_{_4}})|0000\rangle_{r_1r_2r_3r_4}/2,
\end{eqnarray}
The system of four qubit state is a reasonable 
 compromise between a high number of  entangled states and 
resource available for  numerical  computations.
The evolution of the total system can be obtained using  Eq. (\ref{fstate}), so that
a  typical form of the following term evolves as
\begin{equation}
\ket{1}_{e_{_1}}\ket{0}_{e_{_2}}\ket{0}_{e_{_3}}\ket{0}_{e_{_4}}
 \ket{0}_{r_{_1}} \ket{0}_{r_{_2}} \ket{0}_{r_{_3}}\ket{0}_{r_{_4}}\longrightarrow
 \left[u_1(t)\ket{1}_{e_{_1}} \ket{0}_{r_{_1}}+
v_1(t)\ket{0}_{e_{_1}}\ket{1}_{r_{_1}}\right ]
\;\ket{0}_{e_{_2}} \ket{0}_{e_{_3}}\ket{0}_{e_{_4}} \ket{0}_{r_{_2}}\ket{0}_{r_{_3}}\ket{0}_{r_{_4}},  
\label{exampd}
\end{equation}
where $u_1(t)$ and $v_1(t)$ (see  Eqs.(\ref{fstate}), (\ref{ana})) are interaction parameters
associated with the first excitonic qubit which we label via the subscript $1$.
Continuing with the same manner with the remaining three  excitonic qubits, we obtain
\begin{eqnarray}
\label{evolveag}
\nonumber  \left| {\phi _t } \right\rangle&=&  [\left|{000}\right\rangle _{e_{_1}e_{_2}e_{_3}}
\left| { 0 0 0}\right\rangle _{r_{_1}r_{_2}r_{_3}}(u_4(t)|1\rangle_{e_{_4}}|{0}\rangle_{r_4}+
v_4(t)|0\rangle_{e_{_4}}|1\rangle_{r_4})\\
\nonumber ~&~&+ \left| {000}\right\rangle _{e_{_1}e_{_2}e_{_4}} \left|
{ 0 0 0}
\right\rangle _{r_{_1}r_{_2}r_{_4}}(u_3(t)|1\rangle_{e_{_3}}|{0}\rangle_{r_3})+v_3(t)|0\rangle_{e_{_3}}|1\rangle_{r_3}) \\
\nonumber  ~ &~& +\left| {000} \right\rangle _{e_{_1}e_{_3}e_{_4}} \left|
{ 0 0 0} \right\rangle _{r_{_1}r_{_3}r_{_4}}
   (u_2(t)|1\rangle_{e_{_2}}|{0}\rangle_{r_2})+v_2|0\rangle_{e_{_2}}|1\rangle_{r_2})\\
   ~&~&+\left| {000} \right\rangle _{e_{_2}e_{_3}e_{_4}} \left| { 0 0 0}
   \right\rangle _{r_{_2}r_{_3}r_{_4}} (u_1(t)|1\rangle_{e_{_1}}|{0}\rangle_{r_1}+v_1(t)|0\rangle_{e_{_1}}|1\rangle_{r_1})
  ]/2.
\end{eqnarray}
The reduced density operator of the excitonic qubit subsystem $\rho _{e} (t)
= \rm{Tr}\it _{r} (\left| {\phi _t } \right\rangle \left\langle
{\phi _t } \right|)$ and the reduced density operator of reservoir
subsystem $\rho _{r} (t) = \rm{Tr}\it_ {e} (\left| {\phi _t }
\right\rangle \left\langle {\phi _t } \right|)$ can be obtained, provided details of each individual
system-reservoir parameters, $u_i,v_i$ (i=1-4) are known. The detailed derivation of
the density operator matrices for such  generalized exciton-phonon reservoir 
systems is complex and for ease in numerical evaluations, we consider the simple model
in which  $u_1=u_2=u_3=u_4=u$.  Thus each of the exciton is assumed to be coupled to a independent phonon bath, with  same spectral density. This allows convenient evaluation of 
the entanglement measure of the excitonic qubit (denoted by $E_e$) and phonon reservoir $E_r$ systems,
using Eqs.(\ref{ent}) and  (\ref{nnega}), and at the same time is not  a significant compromise on the 
accuracy of the evaluated results. Three parameters which appear
in the spectral density relation in Eq.(\ref{spectral}) are expected to play critical role in the analysis 
of the entanglement dynamics, (a) time ($t$),
(b)   $\gamma_0$ which is associated with   the exciton relaxation time  and (c)
$\delta$ which is the amount by which the 
 central peak is detuned from the exciton transition frequency.

\begin{figure}[htp]
  \begin{center}
\subfigure{\label{aa}\includegraphics[width=5.5cm]{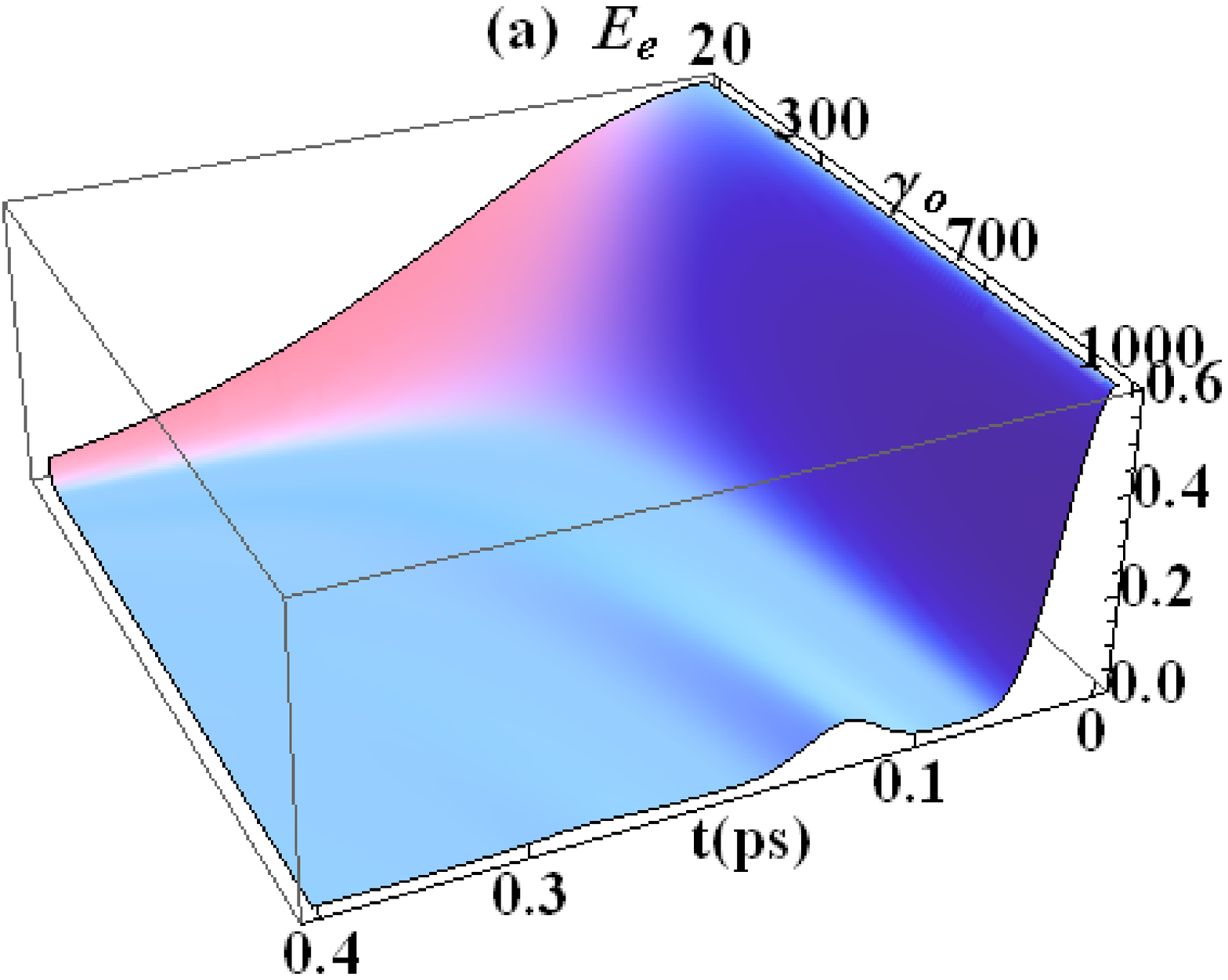}}\vspace{-1.1mm} \hspace{1.1mm}
\subfigure{\label{bb}\includegraphics[width=5.5cm]{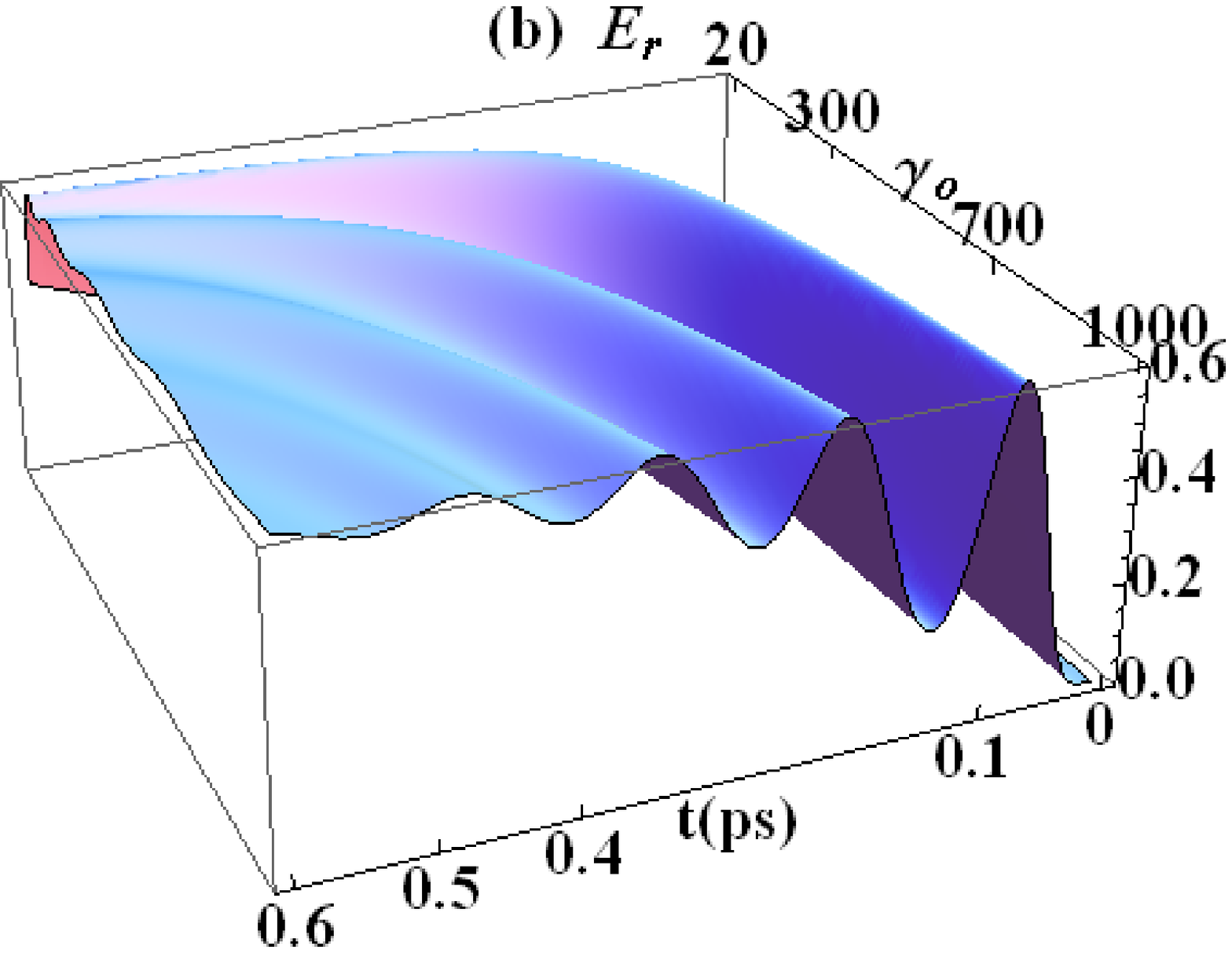}}\vspace{-1.1mm} \hspace{1.1mm}
\subfigure{\label{cc}\includegraphics[width=5.5cm]{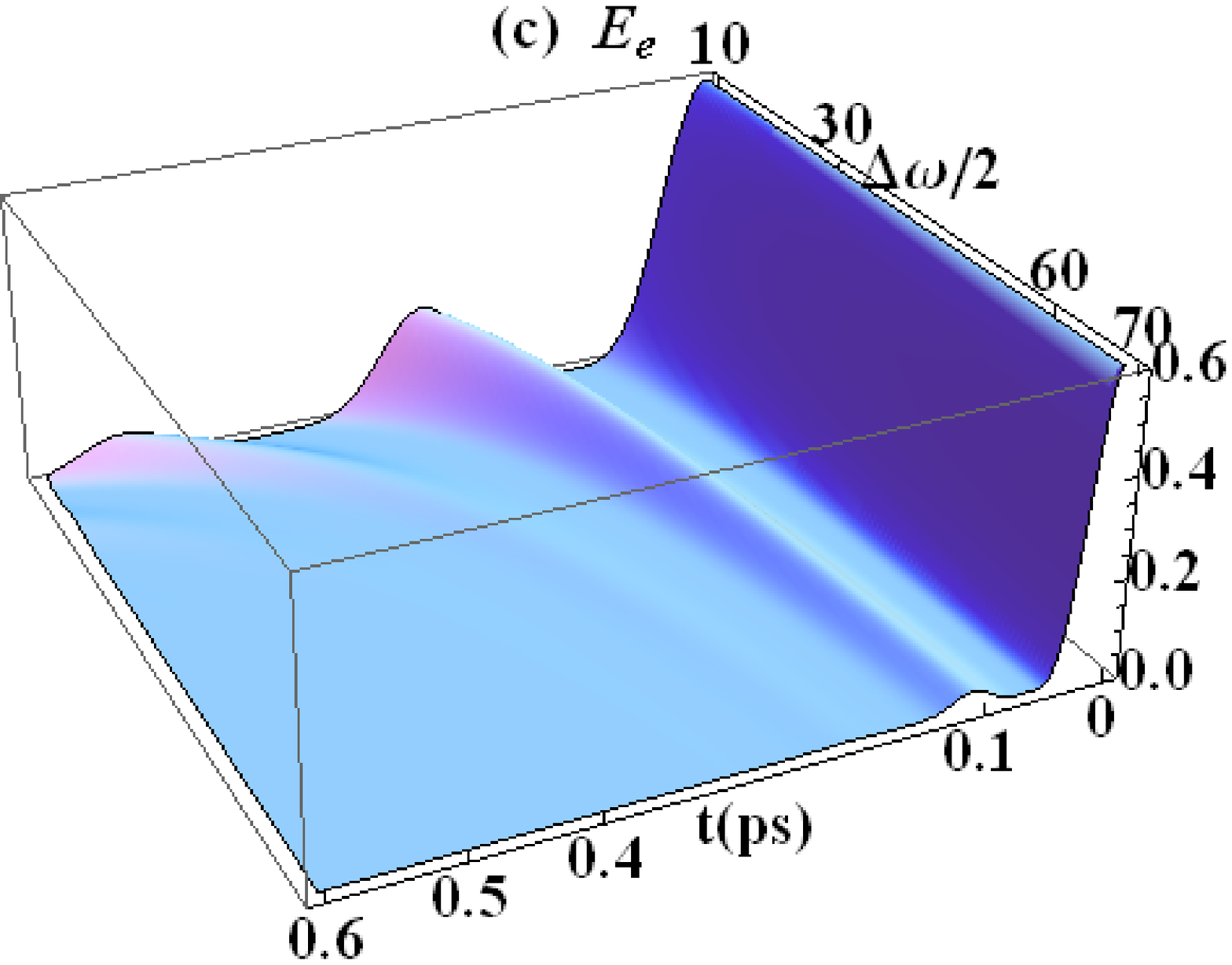}}\vspace{-1.1mm} \hspace{1.1mm}
     \end{center}
  \caption{ (a) Entanglement measure of  the  excitonic qubit subsystem, $E_e$
 as function of  time $t$ (ps) and $\gamma_0$ (cm$^{-1}$), evaluated  for 
the $W$ state (Eq.(\ref{wini}) using  Eqs.(\ref{ent}) and  (\ref{nnega}).
The detuning parameter, $\delta$ =0. $\Delta \omega$=80 cm$^{-1}$ \cite{lorenExpt} in 
 Eq.(\ref{spectral}). The parameter estimates are typical of the FMO complex of  \textit{P. aestuarii}  
\cite{lorenExpt} \quad (b) Same as in (a) except that the entanglement measure of  the system  of 
the collective reservoir states $E_r$ is obtained as a 
function of $t$ (ps) and $\gamma_0$ (cm$^{-1}$) \quad (c) 
Entanglement measure of  the  excitonic qubit subsystem, $E_e$
 as function of  time $t$ (ps) and  $\Delta \omega/2$ (cm$^{-1}$), with 
the detuning parameter, $\delta$ =0 and $\gamma_0$=  1000 cm$^{-1}$.
}
\label{wstate}
\end{figure}

We consider both the Markovian ($\Delta \omega  \gg \gamma_0$) and  non-Markovian regimes 
($\Delta \omega  \ll \gamma_0$) as well as the non- resonant case for which $\delta \neq 0$. 
 Fig.~\ref{wstate}a,b illustrates the  evolution of the
entanglement measure of  the  excitonic qubit subsystem (reservoir subsystem), $E_e$  ($E_r$)
 as function of  time $t$ (in picoseconds) and $\gamma_0$ (in units of cm$^{-1}$). The entanglement
measures are evaluated  for  the $W$ state (Eq.(\ref{wini}) using Eqs.(\ref{ent}), (\ref{nnega}) and  (\ref{ana}),
and setting the detuning parameter  at $\delta$ =0. It can be seen that
as $\gamma_0$ increases, the sub-systems move into  the non-Markovian
regimes, and increased oscillations in revivals of entanglement, associated with comparatively  long phonon 
reservoir correlation times, can be seen in both sub-systems. The non-Markovian features are expected
due to inflow of quantum information from the reservoir subsystem back into the excitonic
qubit subsystem. Overall there is net flow of entanglement from the excitonic
subsystems to the phonon reservoir systems after 0.5 ps.  Fig.~\ref{wstate}c also  illustrates
 that for smaller  $\Delta \omega/2 \sim$20 cm$^{-1}$ where there is higher
reservoir correlation time, there is a pronounced revival in the entanglement measure $E_e$. 

\begin{figure}[htp]
  \begin{center}
\subfigure{\label{aa}\includegraphics[width=6.4cm]{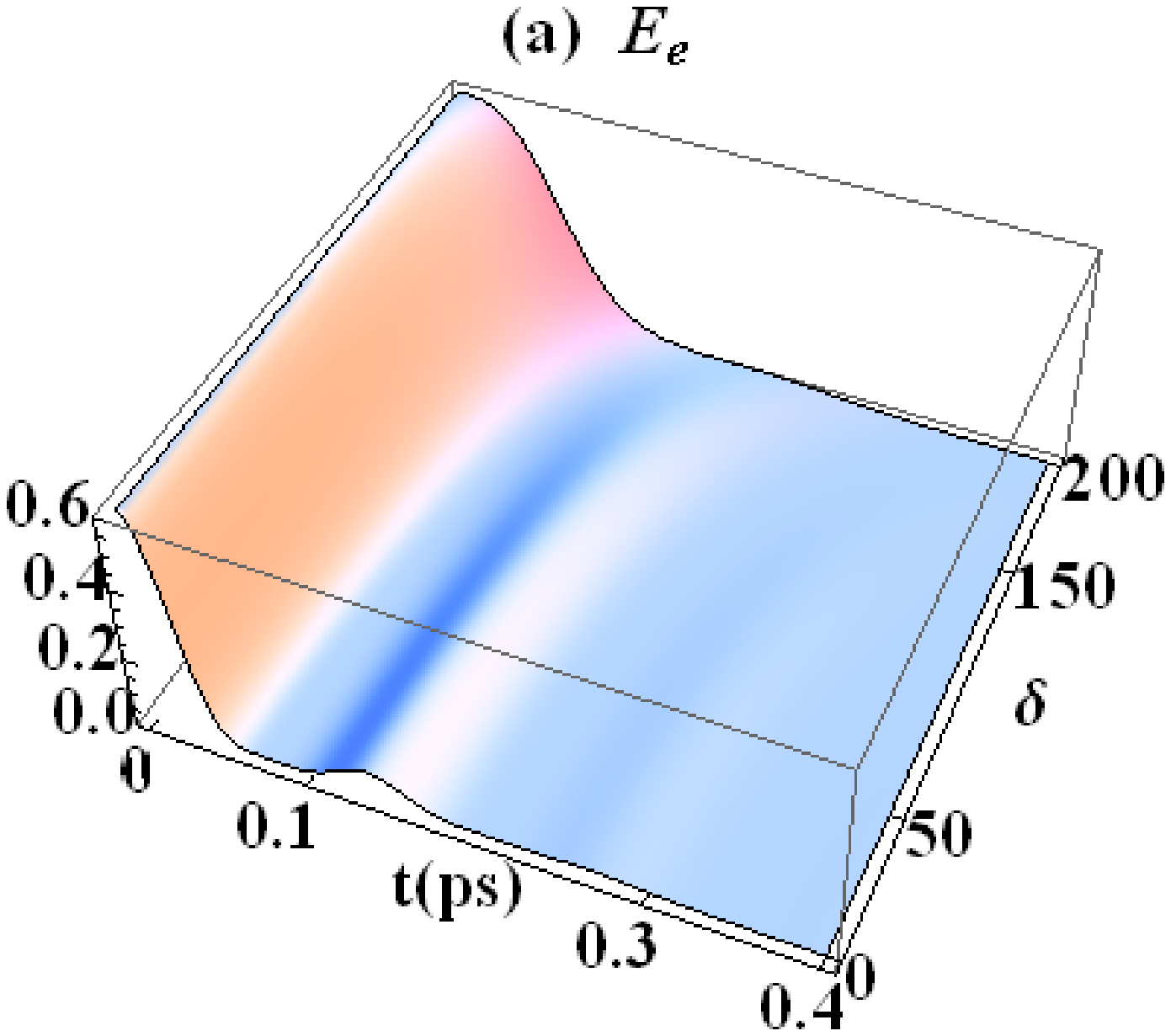}}\vspace{-1.1mm} \hspace{1.1mm}
\subfigure{\label{bb}\includegraphics[width=6.4cm]{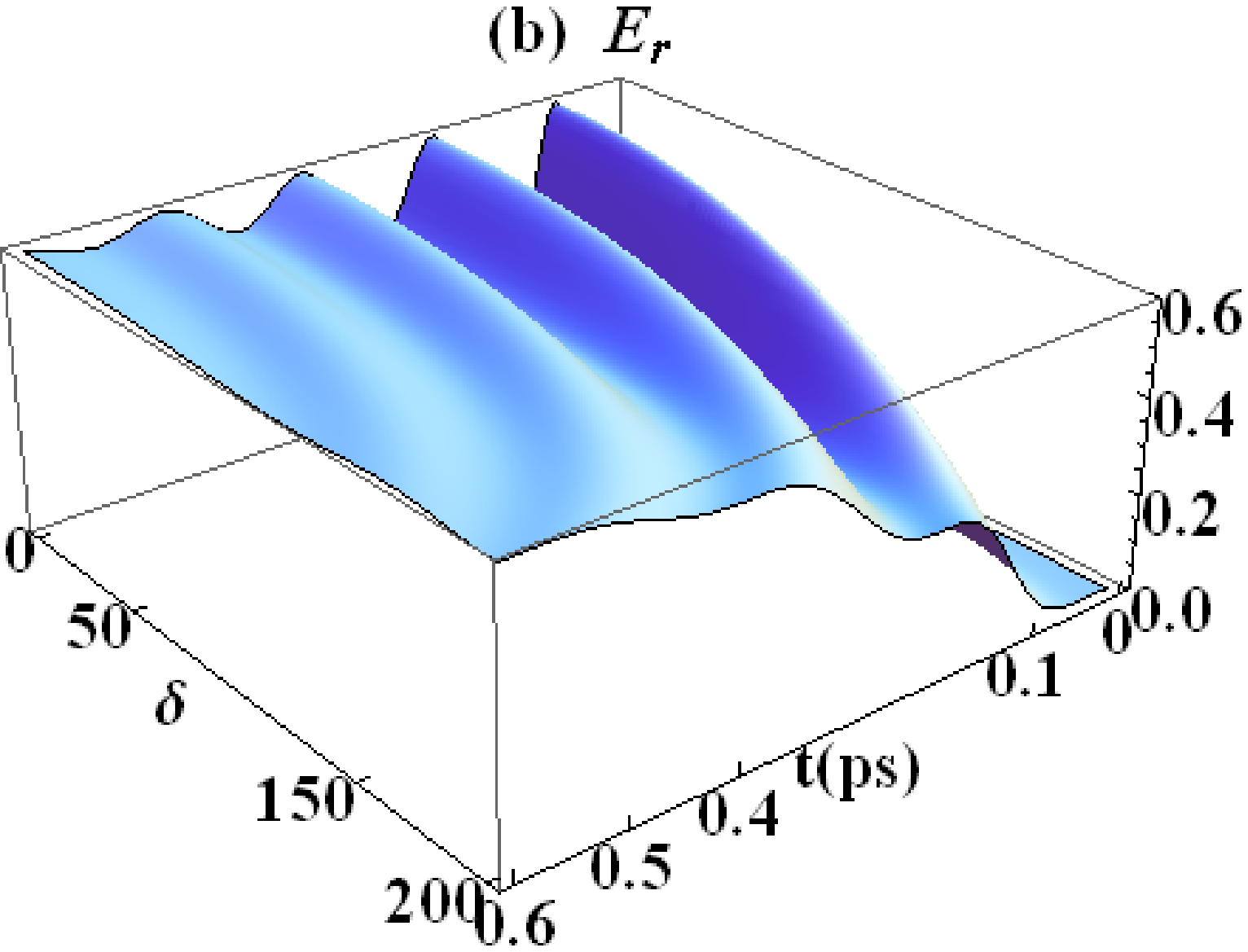}}\vspace{-1.1mm} \hspace{1.1mm}
     \end{center}
  \caption{(a) Entanglement measure of  the  excitonic qubit subsystem, $E_e$
 as function of  time $t$ (ps) and $\delta$ (cm$^{-1}$), evaluated  for 
the $W$ state (Eq.(\ref{wini}) using  Eqs.(\ref{ent}) and  (\ref{nnega}).
 $\gamma_0$ is set at 1000 cm$^{-1}$ and $\Delta \omega/2$=40 cm$^{-1}$.
\\
(b) Same as in (a) except that the entanglement measure of  the system  of 
the collective reservoir states $E_r$ is obtained as a 
function of $t$ (ps) and $\delta$ (cm$^{-1}$)}
\label{dstate}
\end{figure}

Figs. ~\ref{dstate}a,b also show that both the entanglement measures, $E_e$ and $E_r$ decrease
 when the detuning $\delta$ increases, which has been  attributed to the effective decrease in 
coupling between the two interacting subsystems   \cite{chinMar}.
The time duration for  which significant entanglement exists
($\approx$ 0.1-0.3 ps) obtained here is less than the decoherence times noted 
for the  FMO complex of  \textit{P. aestuarii}  \cite{lorenExpt}.
The evaluation of the entanglement measures,  $E_a$
 and $E_r$ based on $GHZ$ states are expected to yield similar qualitative  variations
with respect to change in time $t$, $\gamma_0$ and $\delta$ as  obtained for the $W$ states.
However some quantitative differences due to  the different degree of robustness of the
different multipartite states, can be expected in  the entanglement
measures,  $E_a$  and $E_r$.

It is important to note that the results obtained here are  based on simplified
model systems with underlying assumptions, such as the requirement that
all  qubits  in different Hilbert subspaces experience similar isotropic interactions with the
phonon reservoir subsystem. The qubits are assumed to be independent of each other, and coupled to uncorrelated reservoirs. In the realistic situation, each excitonic qubit may share  one or more
 BChl sites as shown in Table  \ref{tblqu}, which means that the reservoir subsystems are correlated as
well. Due to the large energy difference between each qubit system (about 100 cm$^{-1}$, Table  \ref{tblqu}),
the energy profile of phonon modes linked to different qubit  are expected to be different,
which to some extent justifies the assumption that the reservoir system are uncorrelated.
The basic models employed here also circumvent the inclusion of more realistic features such
the correlated fluctuations of the exciton  transition energies.
It has been shown that environmental changes lead to correlated fluctuations of the exciton  transition 
and interaction energies with impact on population transfer rates, decoherence rates and 
the efficiency of photosynthetic complexes \cite{van}. Even  interaction 
fluctuations of a small magnitude results in an increase in efficiency of the FMO complex \cite{van}.
Nevertheless,  the use of the assumptions  mentioned earlier have enabled the numerical evaluation of some estimates
of the timescales involved in the entanglement dynamics of  the system under study,
and to help understand  on a qualitative level  some critical parameters
responsible for  quantum coherence  in light-harvesting systems. 
The incorporation of realistic features will lead to slight changes in the magnitude
of the  entanglement measures considered here and in the next Section, however the qualitative
features are expected to be retained.

\section{Meyer-Wallach  measure of excitonic qubits in the FMO complex}\label{MWfmo}

In this section, we evaluate  the Meyer-Wallach  measure 
for a system of   photosynthetic  qubits,  with each qubit  interacting with its own 
reservoir of phonons. This monotone measure was defined  by Meyer and Wallach \cite{meyer}
as a single scalar measure of pure state entanglement for a system of more several qubits.
For the case of   $N$ qubits, the Meyer-Wallach  measure 
 is based on the entanglement of each qubit
  with the remaining $(N-1)$-qubits as follows
\begin{equation}
\label{mw}
Q = \frac{1}{N} \sum_{k=1}^{N} 2(1-{\rm Tr}[\rho_k^2]),
\end{equation}
where $\rho_k$ is the reduced density matrix of the $k$th qubit  obtained 
after tracing out all the remaining qubits. 
The Meyer-Wallach  measure has inherent deficiencies 
in that it is unable to distinguish states which
are fully inseparable from states which are separable 
into states of  subsystems. Nevertheless,  it is a useful qualitative measure in the investigation of
subtle features of entangled systems, and which 
we will utilize to analyze the  dynamics of the
photosynthetic system under study in this Section.

We first consider the joint evolution of a pair of  non-interacting excitonic qubits
coupled to uncorrelated phonon reservoirs, and which are  initially prepared in  the following  state 
\bea
\label{dws} 
|\phi_0\rangle &=& \left [ a (|00\rangle_{e_{_1}e_{_2}}+b |11\rangle{e_{_1}e_{_2}} \right ]
|00\rangle_{r_1r_2},
\end{eqnarray}
The state above  evolves as a four-qubit multipartite state influenced by the real coefficients $a, b$.
There are obvious differences between the entangled states in Eq.(\ref{dws}) and (\ref{wini}). 
The density matrix of the excitonic subsystem derived from the global state in Eq.(\ref{dws}) and expressed
within the basis $(\ket{0 \;0},\ket{0 \;1}\ket{1 \;0}\ket{1 \;1})$ appear as
\begin{equation}\label{mate}
\rho_{e_{_1},e_{_2}}(t)=\left(
\begin{array}{cccc}
 f_1(t) & 0 &0 &f_5(t) \\
0 & f_2(t) &0 &0 \\
0 &0 & f_3(t) &0 \\
  f_5(t) &0& 0 & f_4(t)\\
\end{array}
\right).
\end{equation}.
For $t \ge 0$, the matrix elements of the matrix appearing in  Eq.(\ref{mate})
evolve as 
\begin{eqnarray}
f_1(t)&=& a^2+b^2\; v_1(t)^2 \;v_2(t)^2, \nonumber \\
 f_2(t)&=& b^2\; v_1(t)^2 \;u_2(t)^2, \nonumber \\
f_3(t)&=& b^2\; u_1(t)^2 \; v_2(t)^2, \nonumber \\
 f_4(t)&=&b^2\; u_1(t)^2 \;u_2(t)^2, \nonumber \\ f_5(t)&=& a\; b \; u_1(t)\; u_2(t). 
\nonumber 
\end{eqnarray}
where $u_i,v_i$ are conversion functions associated with the $i$th exciton and 
its phonon reservoir. The problem can be extended to the photosynthetic system
examined earlier by substituting $u$ with the form obtained in 
 Eq.(\ref{ana}). It should be noted that the
 $X$-state structure in Eq.(\ref{mate}) preserve its $X$-form
during evolution. 

For density matrix given in Eq.(\ref{mate}), an explicit expression of the Meyer-Wallach measure
for  the $N$-exciton multipartite state is obtained as \cite{thilazeno,thilapho}
\be
\label{mwQ}
Q = 2 a^2 b^2 + 4 b^2 u^2 v^2
\ee
where it is assumed that  $u_i=u$ for all the $N$ exciton-reservoir interacting system. 
Due to this assumption, the Meyer-Wallach measure, $Q$ in Eq.(\ref{mwQ}) is independent of
$N$. For the elaborate and more realistic  models, in which some exciton-reservoir systems 
evolve differently from adjacent pairs, $Q$ becomes a complicated  function of the  differences in 
system-reservoir dynamics in separate 
Hilbert subspaces. For this reason, we chose a highly simplified scenario of 
the same mode of exciton-reservoir for all $N$ qubits, as was also the case in Sec. \ref{wfmo}.

\begin{figure}[htp]
  \begin{center}
\subfigure{\label{aa}\includegraphics[width=5.6cm]{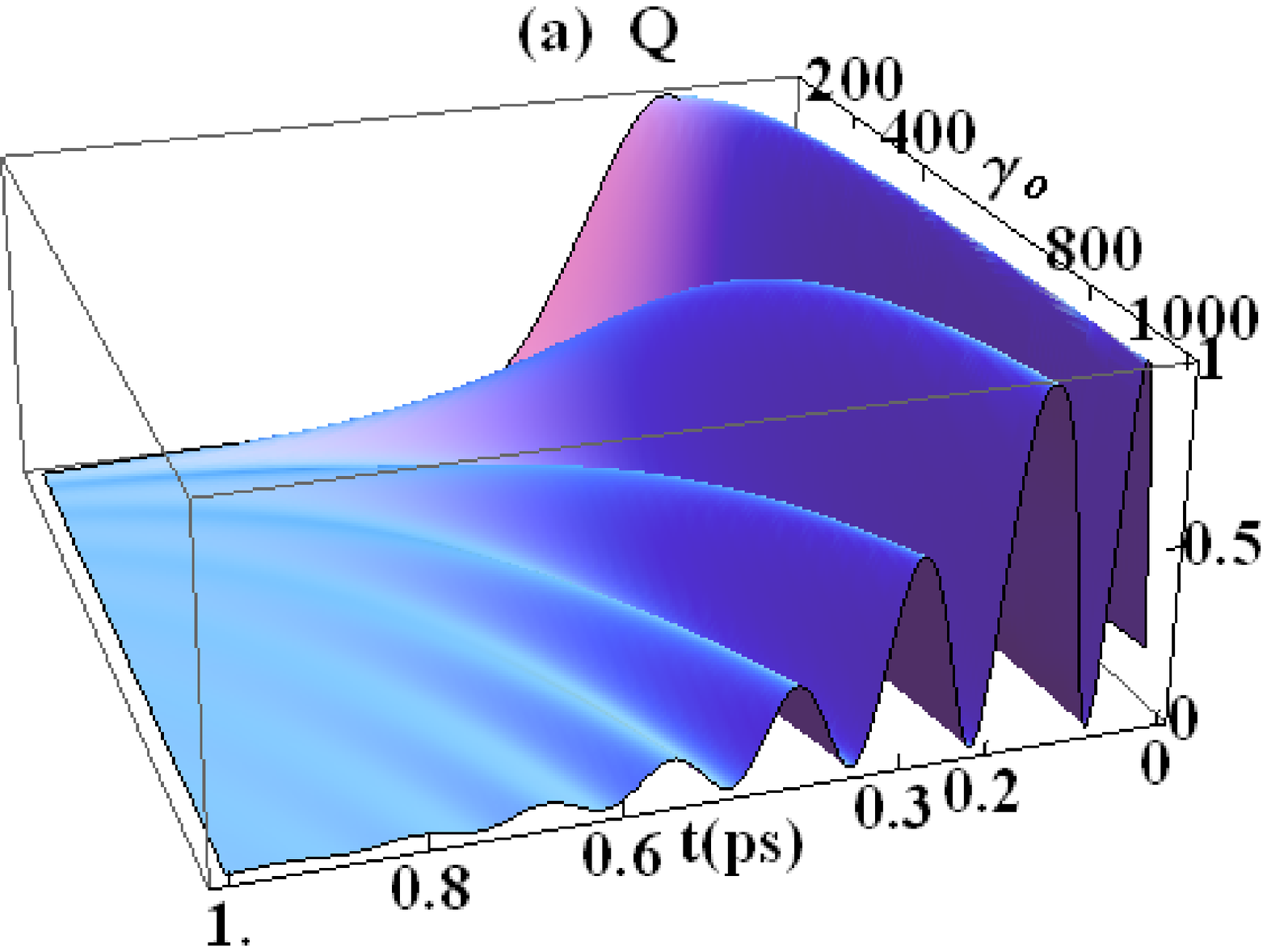}}\vspace{-1.1mm} \hspace{1.1mm}
\subfigure{\label{bb}\includegraphics[width=5.6cm]{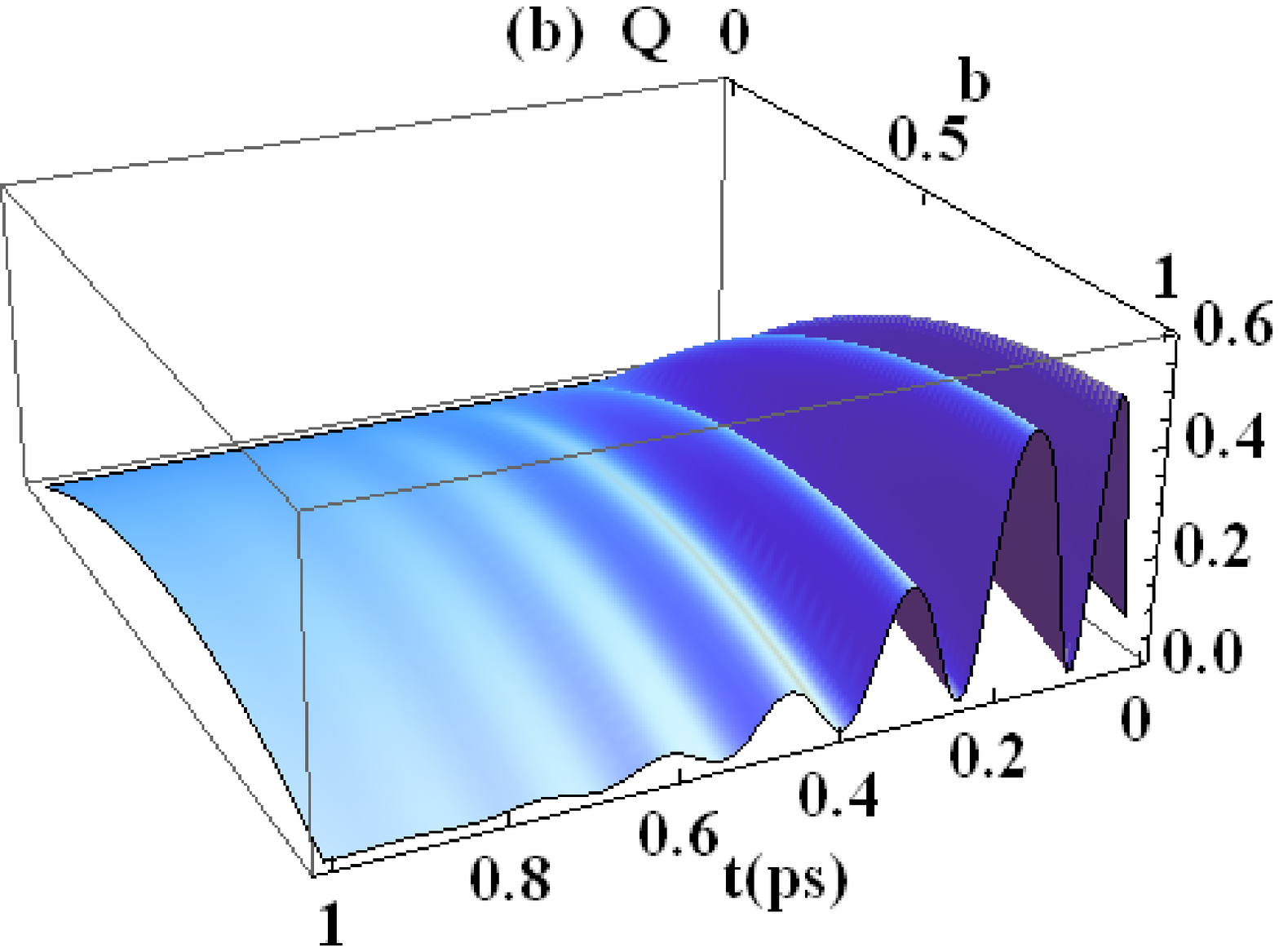}}\vspace{-1.1mm} \hspace{1.1mm}
\subfigure{\label{cc}\includegraphics[width=5.6cm]{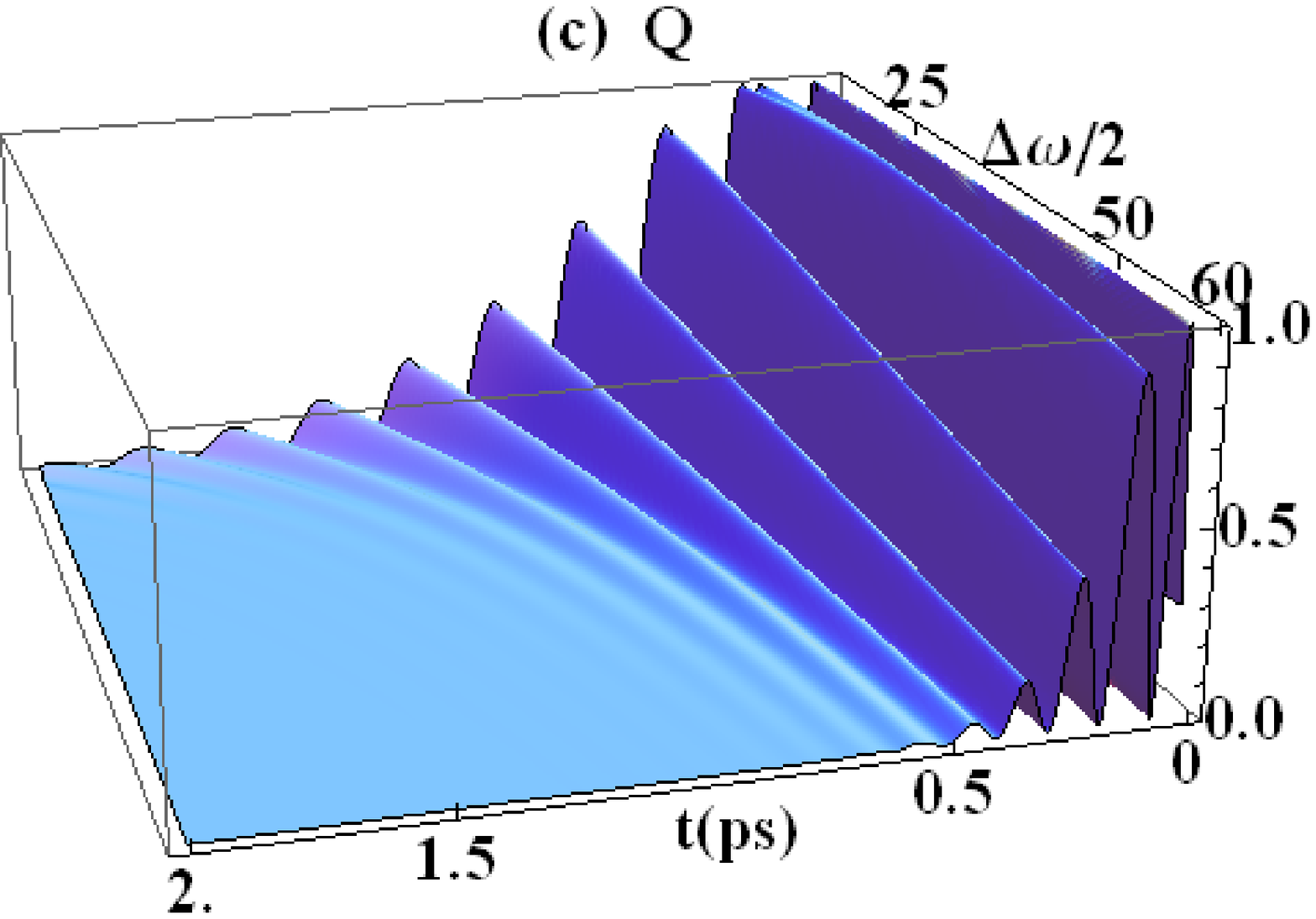}}\vspace{-1.1mm} \hspace{1.1mm}
     \end{center}
  \caption{ (a) Meyer–Wallach measure $Q$ (Eq.(\ref{mwQ})) as a function of  time $t$ (ps) and 
$\gamma_0$ (cm$^{-1}$), evaluated  for  the state in Eq.(\ref{dws}) using  Eq.  (\ref{mwQ}). 
$a$=0, $b$=1 and $\Delta \omega/2$=40 cm$^{-1}$.  \quad (b) Meyer–Wallach measure $Q$ as a function of  time $t$ (ps) and  parameter $b$, $\gamma_0$=800 cm$^{-1}$  and $\Delta \omega/2$=40 cm$^{-1}$.
\quad (c) Meyer Wallach measure $Q$ 
as a function of time $t$ (ps) and $\Delta \omega/2$, $\gamma_0$=1000 cm$^{-1}$, a=0, b=1.
}
\label{Qstate}
\end{figure}

Substituting the form of $u$ given in Eq.(\ref{ana}) into Eq.(\ref{mwQ}),
 the evolution dynamics of the Meyer-Wallach $Q$ measure is plotted in  Fig.~\ref{Qstate}a,b.
The figure shows  expected differences  between the $Q$ measure and the entanglement measures
$E_e$ and $E_r$ obtained in the preceding Section. Unlike $E_e$, the $Q$ measure incorporates the entanglement
of the global system which constitutes the  reservoir subsystems as well.  As a consequence, the 
magnitude of $Q$ shows slower decay with time $t$ at large $\gamma_0 \approx 800 cm^{-1}$ as seen
in  Fig.~\ref{Qstate}a. $Q$ is also sensitive to the correlation parameters $a$ and $b$ (see  Eq.(\ref{dws})) as illustrated in  Fig.~\ref{Qstate}b. As in the case of the entanglement measures
$E_e$ and $E_r$, there is increased revivals in entanglement as the degree of non-Markovian
strength increases.  These revivals can be seen as a signature 
of coherence, and hence the correlation parameter $b$ plays a critical role in the
length of time over which such revivals can be  maintained.
In general, the maximum $Q=1$ is obtained for $a=0,b=1$ with $u_1=u_2=\frac{1}{\sqrt{2}}$.

\section{Teleportation and  splitting fidelities of excitonic qubits in the FMO complex}\label{tele}

In this section, we apply the results of teleportation fidelities
obtained by  Chaves et. al. \cite{chaves} to  analyze the entanglement dynamics of the
FMO complex detailed in Sec. \ref{fmo}. Using  the concept that multipartite entanglement 
is linked to the teleportation  and quantum information splitting protocols \cite{chaves}, we
examine the effect of the
phonon reservoir on  the  entanglement properties of the excitonic qubits in the FMO complex.
This is also linked to the possibility that 
the entangled qubit states in the photosynthetic complex may act as a resource
for the teleportation and quantum information splitting protocols.

 In the teleportation process  \cite{ben93,tele2,pop} , two parties conventionally referred to as 
Alice and Bob  share an entangled bipartite state, with the quality of the teleportation 
measured by the fidelity 
$f=\langle\psi|\rho|\psi\rangle$, where $\rho$ is the state obtained
after teleportation of an
unknown quantum state $\left\vert \psi\right\rangle$.
For  the unknown state of dimension $d$, 
the maximum achievable  fidelity $f_{\max}$ for
a given bipartition of the state used as the medium of transmission is \cite{chaves}
$f_{\max}\leq\frac{2+2\mathcal{N}}{d+1}$ where 
 the negativity $\mathcal{N}$ is associated with the given bipartition.
The bipartite mixed state leads to imperfect teleportation, and thus reduce the
fidelity of teleportation to values below those of 
 the pure states which yield a fidelity of $\frac{2}{3}$ \cite{pop}.

Chaves et. al. \cite{chaves} considered a viable connection between
 global entanglement quantifiers in entangled  systems and  
the robustness of tasks such as teleportation and 
the splitting of quantum information \cite{split1,split2}, which is 
a generalization of the  teleportation protocol.
During the splitting process,  an unknown
qubit $\left\vert \Psi_{0}\right\rangle =\cos\left(
\theta/2\right) \left\vert 0\right\rangle +e^{i\phi}\sin\left(
\theta/2\right)  $ $\left\vert 1\right\rangle $ is sent 
 to $N$ other parties via a shared entangled state.  The no-cloning theorem ensures that 
 only  a single  copy of $\left\vert \Psi_{0}\right\rangle $ can be 
retrieved.  The association between the degree of cooperation 
of the $N$ parties during the  extraction process and 
the  global entanglement of the shared $GHZ$ or $W$-like states under decoherence were
examined in Ref.\cite{chaves}. The results that is of particular relevance in the context
of photosynthetic systems, are the  explicit forms of teleportation and quantum information splitting fidelities
 obtained in Ref.\cite{chaves} for  protocols associated with  
$GHZ$ and $W_{A}$ states, under the amplitude-damping channel.
The $W_{A}$ state is  associated with $N+1$ qubits and of the form
\be
\left\vert W_{A}^{N+1}\right\rangle =\frac{1}{\sqrt{2}}\left[
\left\vert 0\right\rangle\left\vert W_{N}\right\rangle  +\left\vert
1\right\rangle\left\vert 0\right\rangle^{\otimes N} \right]  ,
\label{asW}
\ee
We note the similarities between the  amplitude-damping channel and
 the action of the phonon reservoir
on the photosynthetic qubit as detailed in Sec. \ref{fmo}. Within this channel,
 the population of the excited state is reduced by a factor $1-p$, where 
the $p$ is a  parameterization of time.  For the excitonic qubit interacting with a phonon bath, 
 one has $p=1-|u(t)|^2$,  where u(t) appears in  Eq.  (\ref{ana}).
At initial time $t=0$, $u(t)=1$ and $p=0$, and at
t$\rightarrow\infty$, $u(t)=0$ and $p=1$.

The   teleportation fidelity, averaged over all the possible input states
$\left\vert \Psi_{0} \right\rangle $,  was obtained   as \cite{chaves}
\be 
F_{_{GHZ}} =\frac{1}{6} \left[ 2+\left(  1-p\right)^{N-1}(2-p) \right.
+\left. 2\left(  1-p\right)  ^{N/2}+p^{N-1}(1+p) \right]
\label{fghz}
\ee  
for $GHZ$ resource states. In the case of the $W_{A}$ state associated with $N+1$ qubits, 
$\left\vert W_{A}^{N+1}\right\rangle =\frac{1}{\sqrt{2}}\left[
\left\vert 0\right\rangle\left\vert W_{N}\right\rangle  +\left\vert
1\right\rangle\left\vert 0\right\rangle^{\otimes N} \right]$,
the teleportation fidelity was obtained as \cite{chaves} 
\be 
F_{_{W}} =\frac{1}{3}\left( 3-2p+p^{2} \right)
\label{fw}
\ee
Chaves. et. al. \cite{chaves} likewise obtained simple form 
for the fidelities associated with the 
 complete splitting protocol involving  teleportation followed by the
decodification as
\bea
F_{_{GHZS}} &=& \frac{1}{3}\left[  2-p\left(  1-p\right)  +\left(  1-p\right)  ^{N/2}\right] \label{sg}\\
F_{_{WS}} & = &1-\frac{p}{3}.
\label{sw}
\eea
The full details of the derivation of the fidelities in Eqs. (\ref{fghz}),  (\ref{fw})
 (\ref{sg}) are provided in Ref.\cite{chaves}.

\begin{figure}[htp]
  \begin{center}
\subfigure{\label{aa}\includegraphics[width=5.3cm]{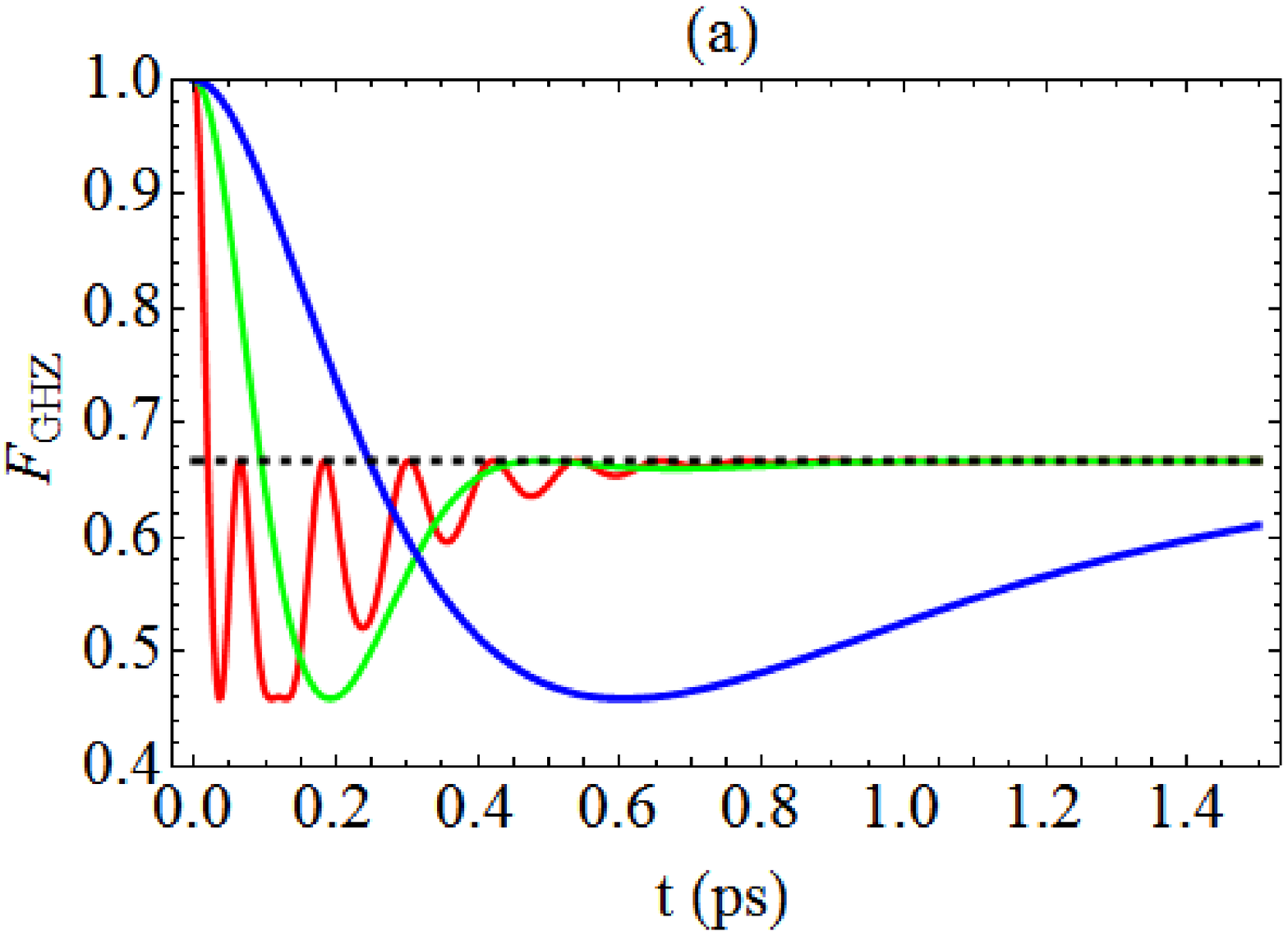}}\vspace{-1.1mm} \hspace{1.1mm}
\subfigure{\label{bb}\includegraphics[width=5.3cm]{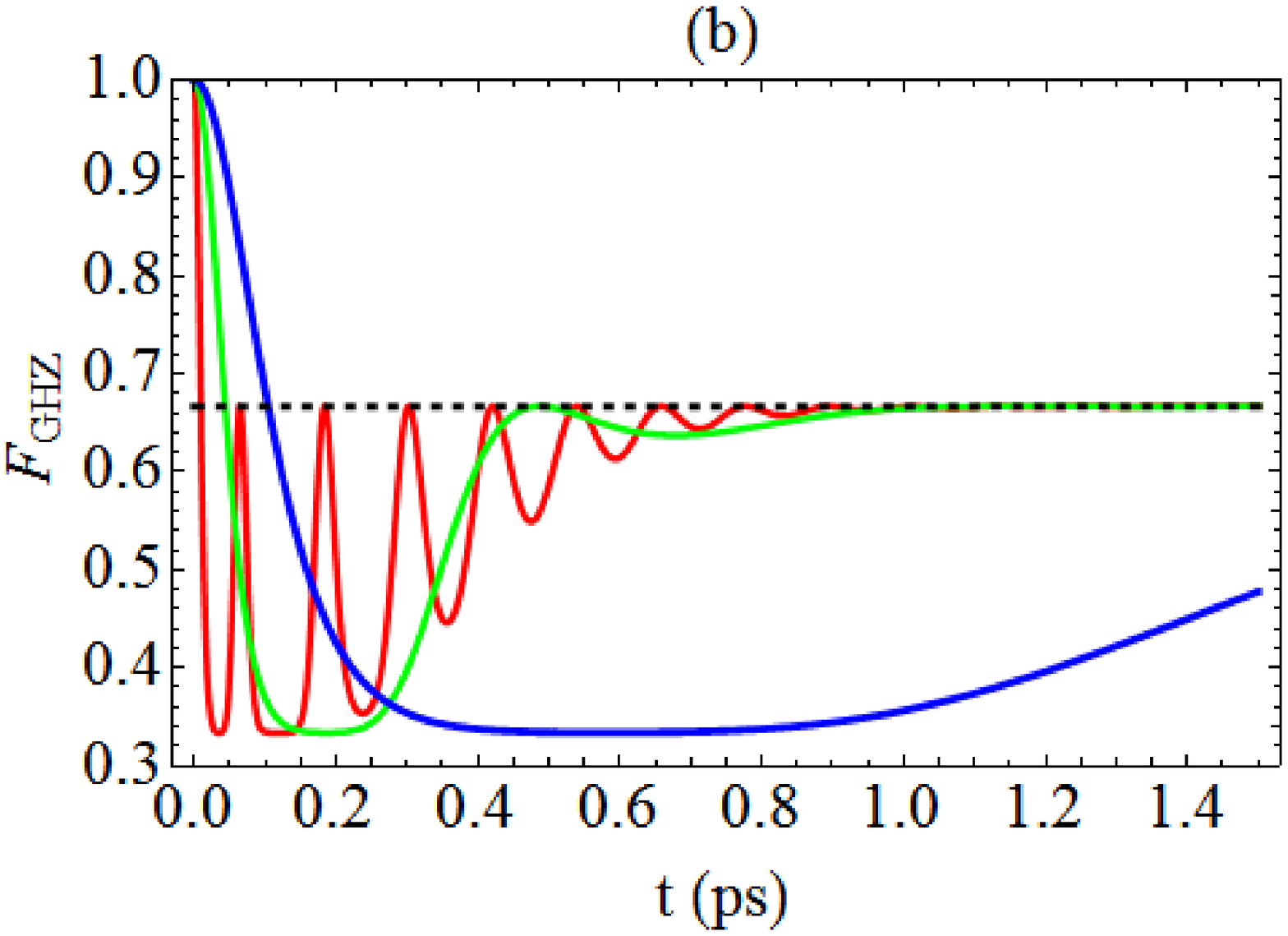}}\vspace{-1.1mm} \hspace{1.1mm}
\subfigure{\label{bb}\includegraphics[width=5.3cm]{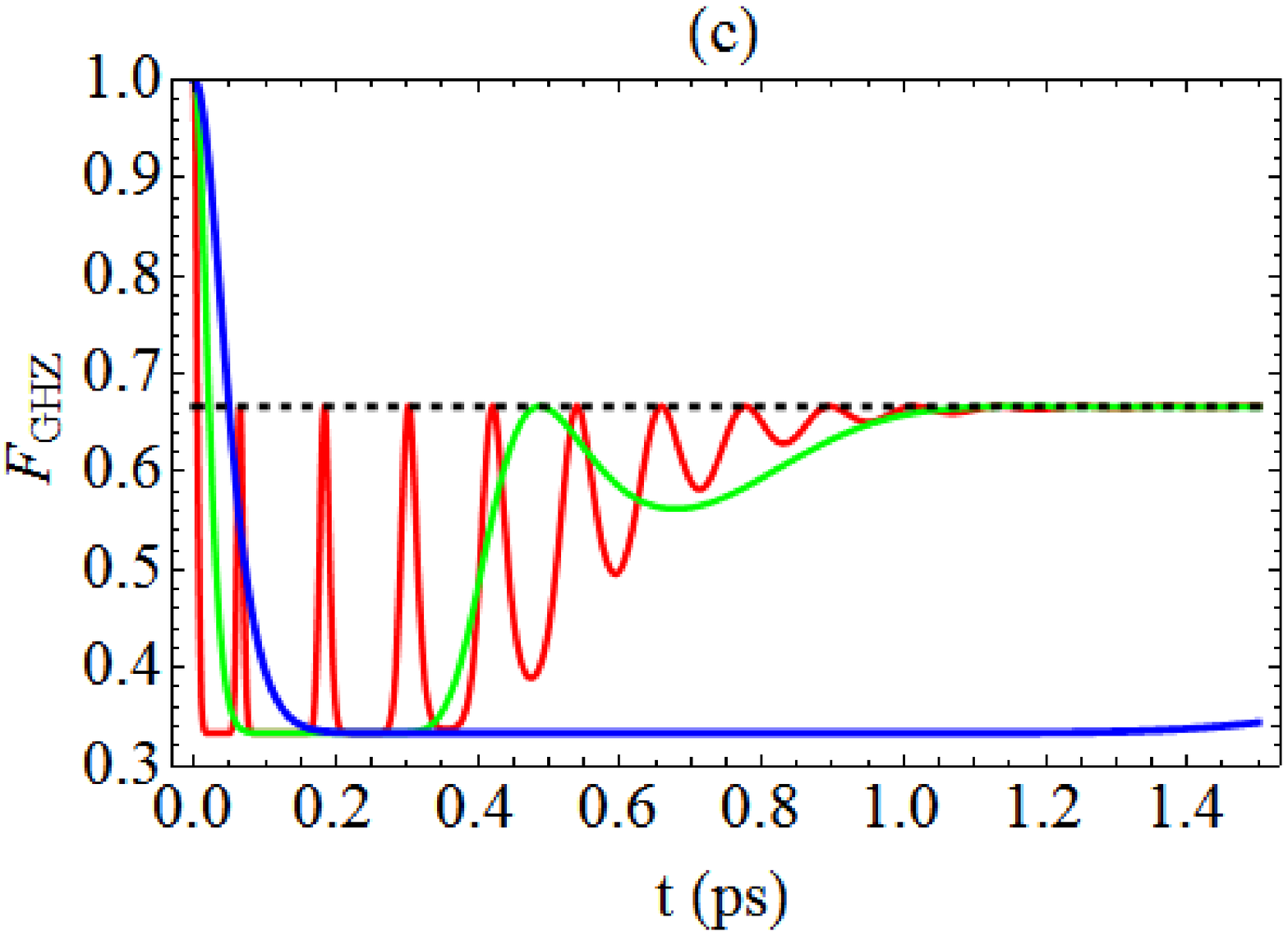}}\vspace{-1.1mm} \hspace{1.1mm}
     \end{center}
  \caption{ (a) Teleportation fidelity, $F_{_{GHZ}}$,  as a function of  time $t$ (ps) and $\gamma_0$ 
[10 cm$^{-1}$ (blue line),  50 cm$^{-1}$ (green line),  1000 cm$^{-1}$ (red line)] based on the  $GHZ$ resource states, using  Eq.  (\ref{fghz}). The detuning parameter  is set at $\delta$ =0 and $N$=4.
$\Delta \omega$=80 cm$^{-1}$ \cite{lorenExpt} in  Eq.(\ref{spectral}). The classical fidelity of $2/3$  is denoted
by dotted lines.  \quad (b)
Same as in (a) except  $N$=16 \quad (c) Same as in (a) except  $N$=64\\
 }
\label{teleg}
\end{figure}

\begin{figure}[htp]
  \begin{center}
\subfigure{\label{aa}\includegraphics[width=5.3cm]{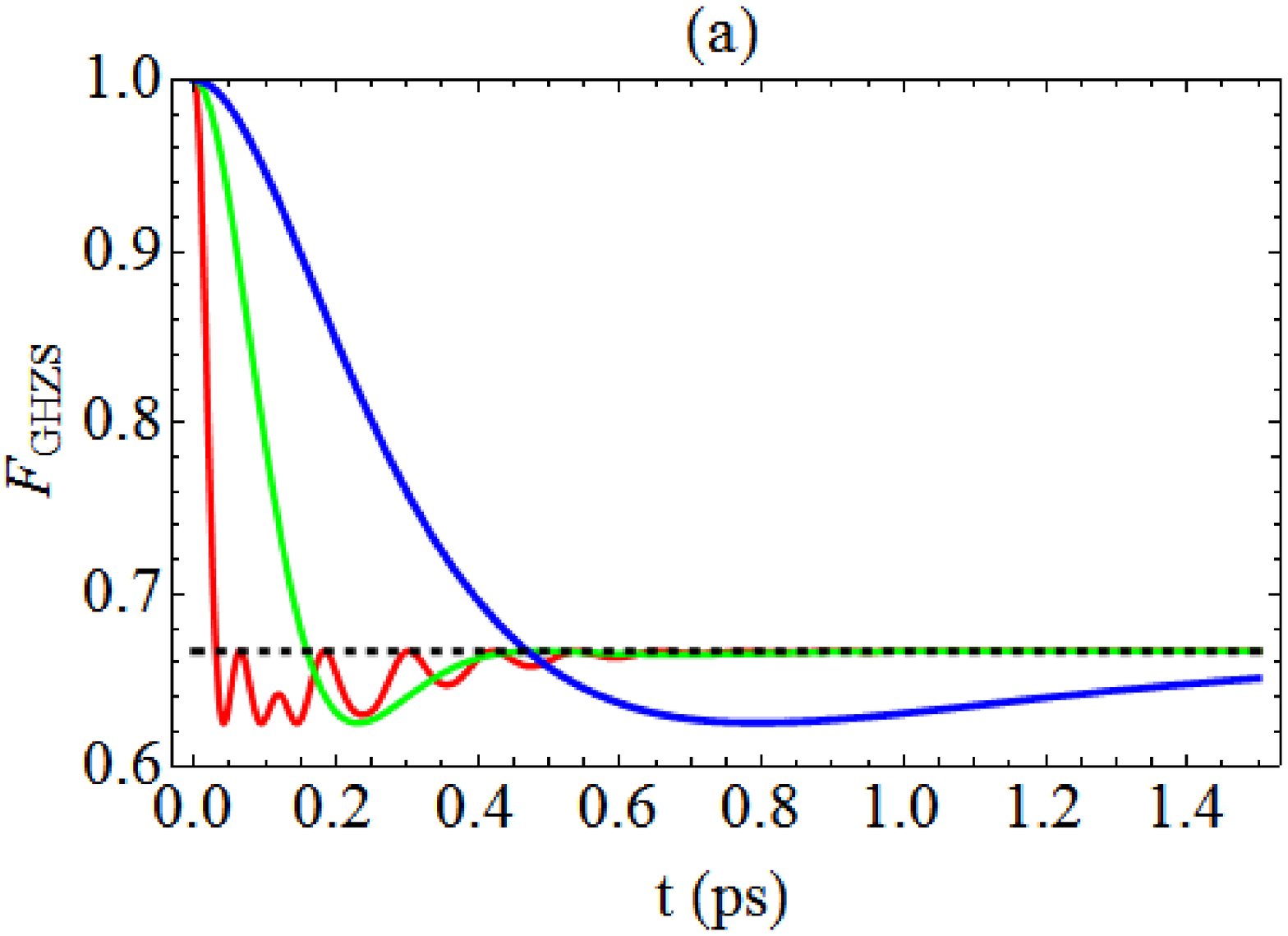}}\vspace{-1.1mm} \hspace{1.1mm}
\subfigure{\label{bb}\includegraphics[width=5.3cm]{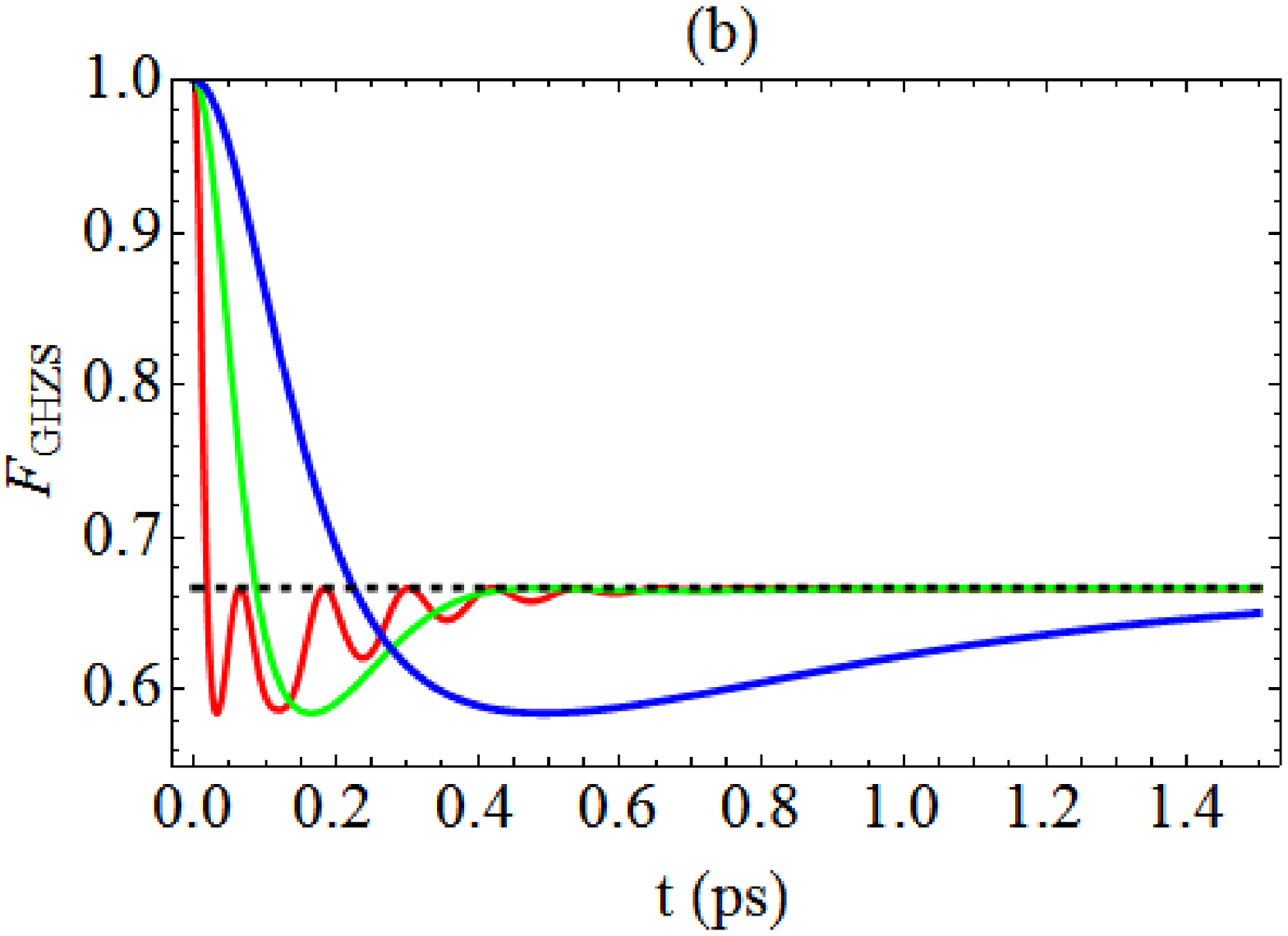}}\vspace{-1.1mm} \hspace{1.1mm}
\subfigure{\label{bb}\includegraphics[width=5.3cm]{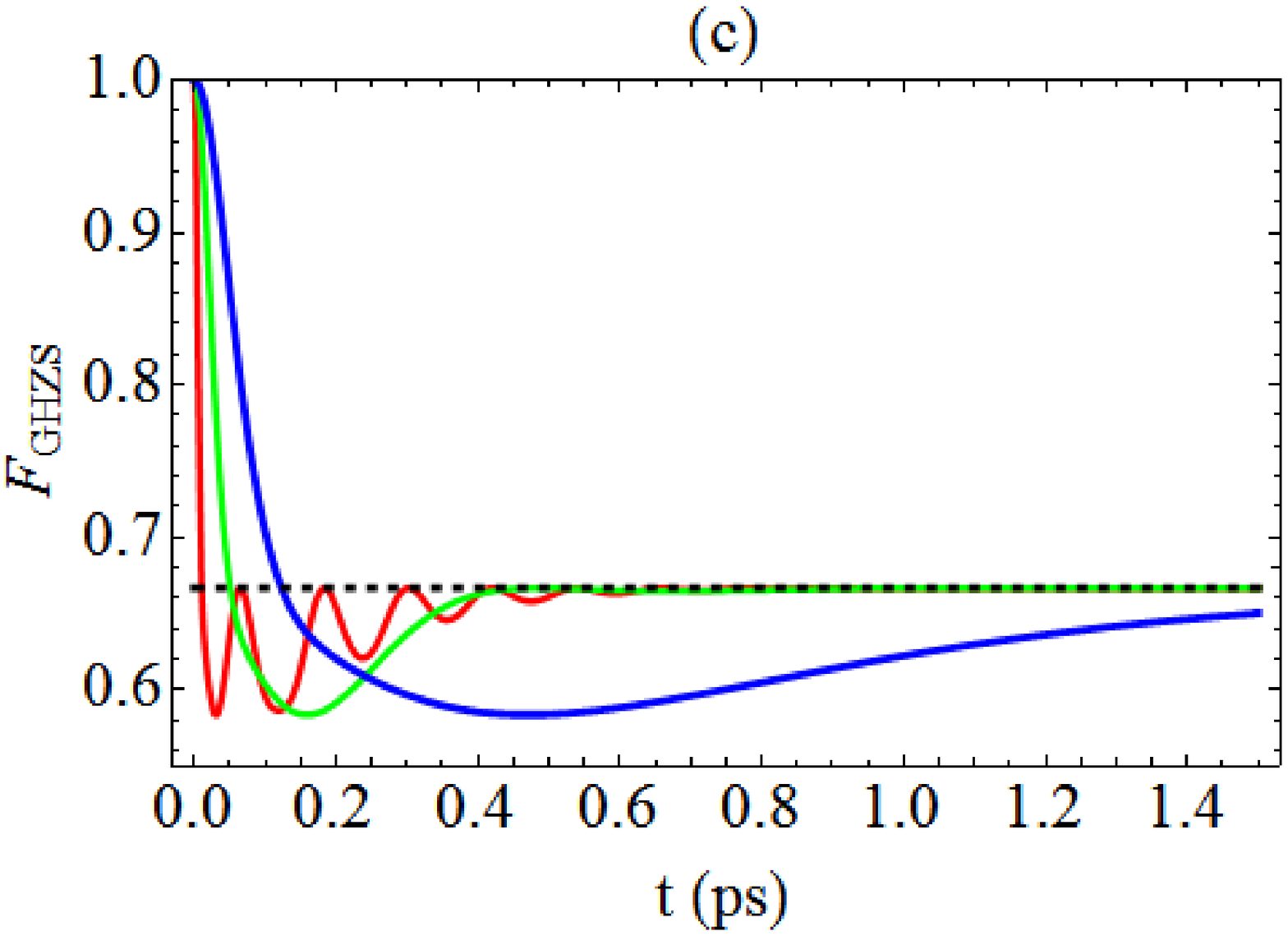}}\vspace{-1.1mm} \hspace{1.1mm}
     \end{center}
  \caption{ (a) Quantum information splitting fidelity, $F_{_{GHZS}}$,  as a function of  time $t$ (ps) and $\gamma_0$ 
[10 cm$^{-1}$ (blue line),  50 cm$^{-1}$ (green line),  1000 cm$^{-1}$ (red line)] based on the  $GHZ$ resource states, using  Eq.  (\ref{sg}). The detuning parameter  is set at $\delta$ =0 and $N$=4.
$\Delta \omega$=80 cm$^{-1}$ \cite{lorenExpt} in  Eq.(\ref{spectral}). The classical fidelity of $2/3$  is denoted
by dotted lines.  \quad (b)
Same as in (a), except $N$=16 \quad (c) Same as in (a) except $N$=64\\
 }
\label{splitg}
\end{figure}

\begin{figure}[htp]
  \begin{center}
\subfigure{\label{aa}\includegraphics[width=7cm]{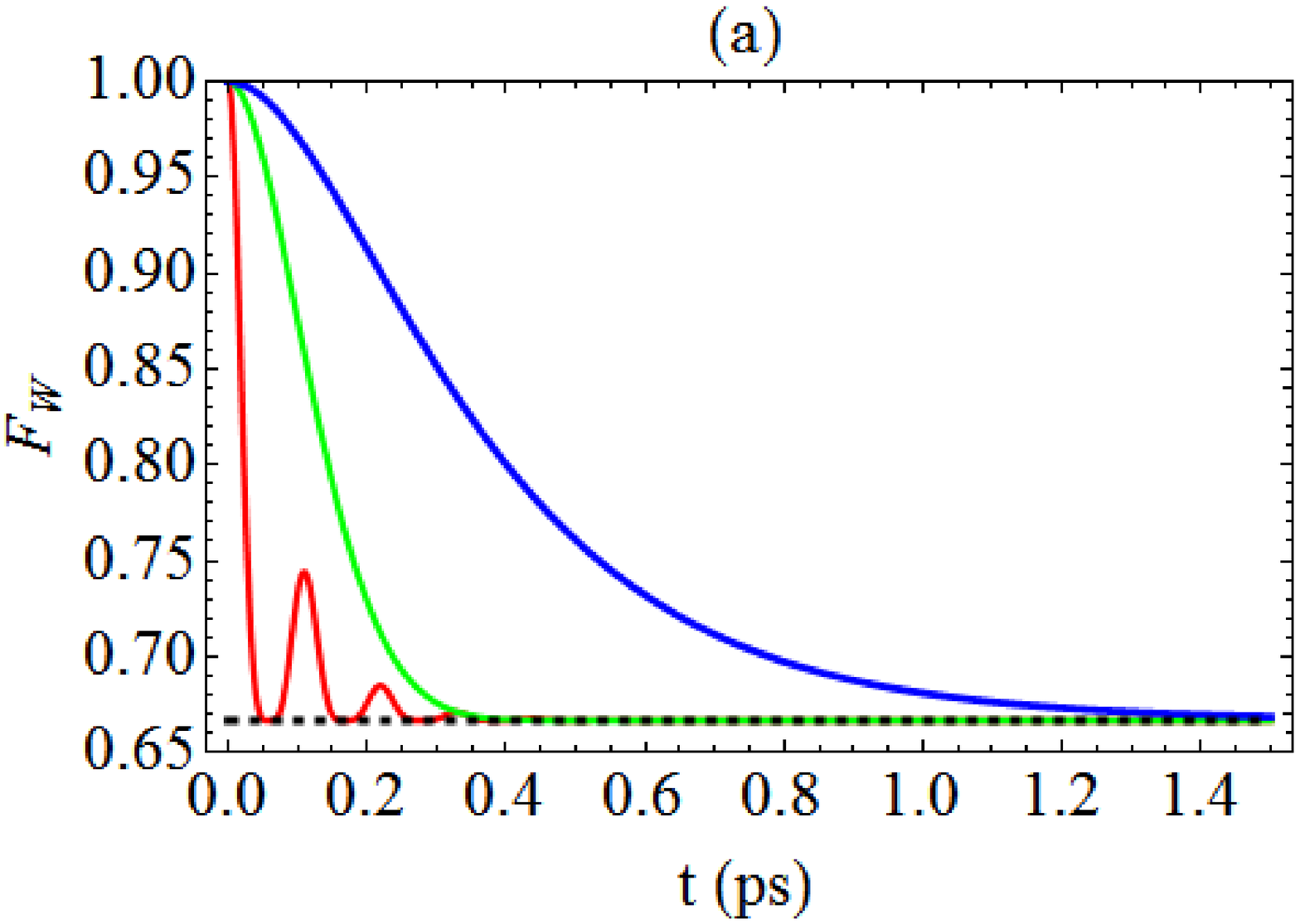}}\vspace{-1.1mm} \hspace{1.1mm}
\subfigure{\label{bb}\includegraphics[width=7cm]{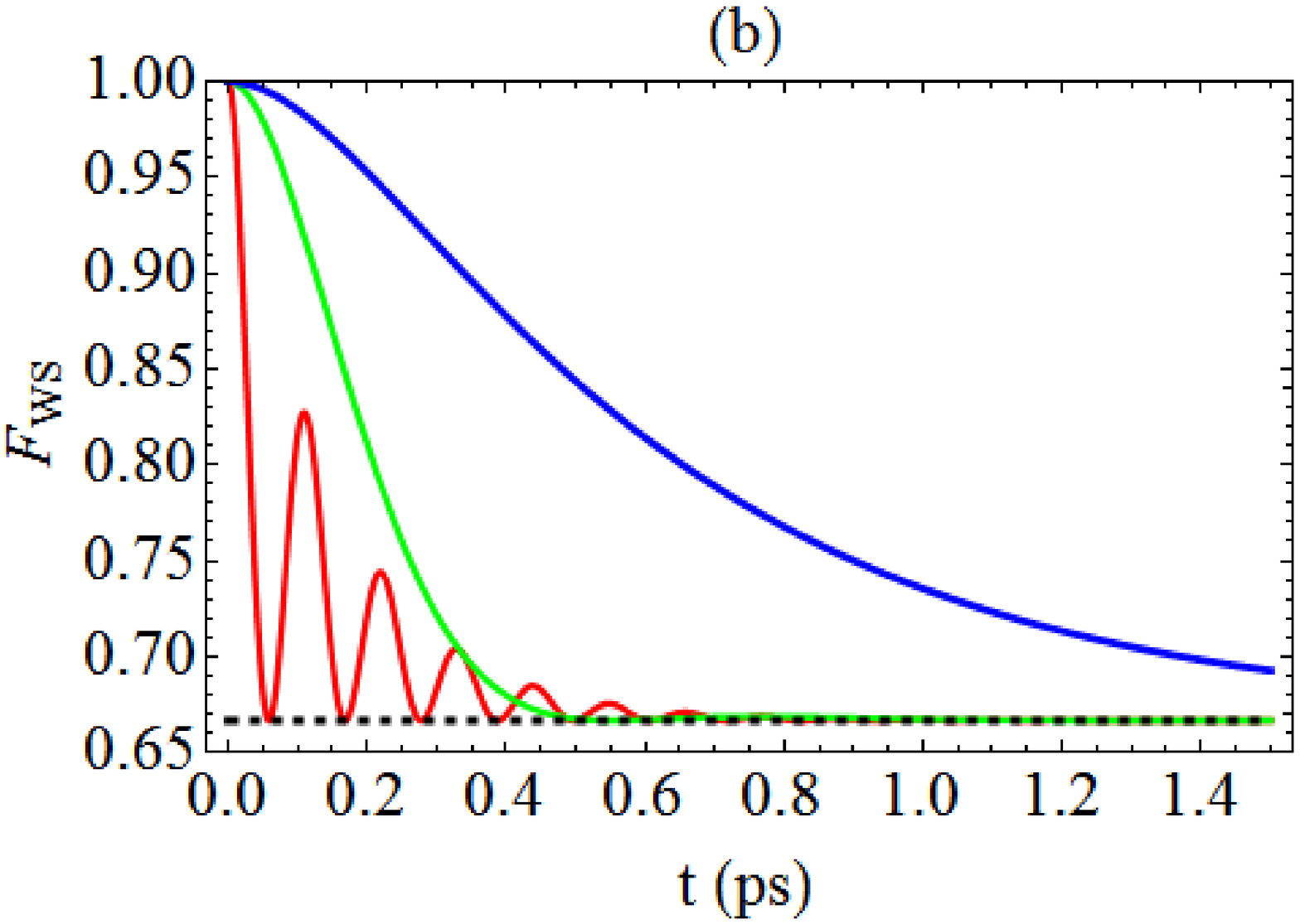}}\vspace{-1.1mm} \hspace{1.1mm}
     \end{center}
  \caption{ (a) Teleportation fidelity,  $F_{_{W}}$,  as a function of  time $t$ (ps) and $\gamma_0$ 
[10 cm$^{-1}$ (blue line),  50 cm$^{-1}$ (green line),  1500 cm$^{-1}$ (red line)] based on the   $W_{A}$ resource states, using  Eq.  (\ref{fw}). The detuning parameter  is set at $\delta$ =0 and 
$\Delta \omega$=80 cm$^{-1}$ \cite{lorenExpt} in  Eq.(\ref{spectral}).  The classical fidelity of $2/3$  is denoted
by dashed lines.
\\
(b)  Quantum information splitting  fidelity, $F_{_{WS}}$,  as a function of  time $t$ (ps) and $\gamma_0$ 
[10 cm$^{-1}$ (blue line),  50 cm$^{-1}$ (green line),  1500 cm$^{-1}$ (red line)] based on the  $W_{A}$ resource states, 
using  Eq.  (\ref{sw}). The detuning parameter  is set at $\delta$ =0 and $\Delta \omega$=80 cm$^{-1}$ 
\cite{lorenExpt} in  Eqs.(\ref{spectral}). }
\label{telew}
\end{figure}

 The evolution dynamics of the  teleportation fidelities for the systems under study here,
 can be obtained as shown in  Fig.~\ref{teleg} and \ref{telew}a by substituting the form 
of $u$ (Eq.(\ref{ana})) into Eqs.(\ref{fghz}) and  (\ref{fw}), 
and using typical parameter estimates of the 
FMO complex of  \textit{P. aestuarii}  \cite{lorenExpt}.
Likewise the fidelities associated with the 
 complete splitting protocol  (teleportation followed by the
decodification) obtained using Eqs.(\ref{ana}), (\ref{sg}) and  (\ref{sw})
are plotted in  Fig.~\ref{splitg} and \ref{telew}b.  
Both  the $GHZ$ and $W_{A}$ resource states
 reach the classical fidelity of $2/3$  in the Markovian and
non-Markovian  regimes, for the two types of  fidelities at  t$\rightarrow\infty$ \cite{chaves}.
The teleportation fidelities associated with the 
$W_{A}$ resource states lie above the classical threshold  fidelity of $2/3$, 
while those of the  $GHZ$ resource states lie below the classical threshold  fidelity.
The results obtained here are in line with
earlier observations by  Chaves et. al.  \cite{chaves} that the
$W_{A}$ states display greater robustness than the  $GHZ$ states.

The   increased revivals in the fidelities with time, for increasing degree of non-Markovian
strength,  can also be noted for both $GHZ$ and $W_{A}$ states, with the amplitude of oscillations  
 bounded either from above or below by the classical threshold value.  These revivals,
appear to last up to time periods of typically 0.5 - 0.7 ps, depending
on  $\gamma_0$,  which is associated with   the exciton relaxation time.
The oscillations in the fidelity measure is indicative of  strong interactions
between the excitonic qubit and the phonon reservoir system.
Fig.~\ref{teleg}a,b,c shows the overall decrease  of the teleportation fidelities with  the number  of qubits $N$ in the $GHZ$ state, however  there is also an associated  increased  oscillations when $N$ is increased at the non-Markovian regime. Unlike the teleportation fidelities,
Fig.~\ref{splitg}a,b,c shows that the splitting fidelity, $F_{_{GHZS}}$
is more robust to changes in the  number  of qubits $N$ in the 
$GHZ$ state.  Comparison of the results in Figs.~\ref{teleg}, \ref{splitg}  and  \ref{telew} 
show  that in general, higher fidelities are obtained in the case of the 
complete splitting protocol   involving teleportation followed by the 
decodification. In particular, there is notable increased oscillations of the fidelities
associated with the complete splitting protocol in the non-Markovian regime
when  $\gamma_0$ is increased to  1500 cm$^{-1}$ (see Fig.  \ref{telew}). 

Any connection between  coherence and fidelity  appears to be subtle,
and a rigorous link between the two measures have been 
not demonstrated so far, and is beyond the scope of this work. 
As the fidelity characterizes the quality of information transmission
through the appropriate quantum channels (e.g. excitonic qubit channels of the FMO complex),
it would not be naive to link  higher fidelities with more robust coherence phenomenon.
Using this reasoning, we can attribute coherent oscillations at physiological temperatures
to quantum information processing tasks involving teleportation followed by the
decodification process involving $W_A $states of the FMO complex. Such tasks in particular,
may become more important at large $\gamma_0$, which is more 
likely to occur at higher temperatures. 

It well known that  one antenna complex serves many FMO complexes, and 
multiple FMOs are connected to a given reaction center, a structural arrangement  which allow
photosynthetic organisms to most effectively organize cellular resources \cite{p1,p3,sener}.
Even though the inter-monomer coupling strengths
are about one order of magnitude smaller than intra-monomer couplings \cite{reng},
the three weakly coupled subunits in the FMO  photosynthetic complex trimer, 
can form a larger cluster of excitonic qubits.
There also exists the occurrence of massive entangled excitonic qubits of $N > 64$ 
across an entire photosynthetic membrane constituting thousands of bacteriochorophylls
as determined using atomic force microscopy in Ref. \cite{sener}. 

In summary, there are  increased revivals in fidelities in the non-Markovian
regime for  resource states ($GHZ$ states, $W$-like states) in contact
with a decoherent environment. This can be attributed to the  comparatively  long phonon 
reservoir correlation times associated with non-Markovian interactions.
Unlike the $GHZ$ states, the $W$-like states are expected to provide greater resilience
to environmental related decoherence during photosynthetic processes. The contribution
to global entanglement due to teleportation followed by the
decodification tasks in $W_{A}$ states  of the FMO complex appear to play a domineering
role during  quantum oscillations at physiological temperatures. 
The possibility of the latter processes accounting for 
 experimental results \cite{pani,collini} which show that coherent oscillations persist up to 1 ps  at physiological temperatures  cannot be ruled out.  These results have important implications
for multipartite states  present in the FMO complex of light-harvesting systems.

\section{\label{con} Conclusion}
 
In this work, we  examined the exciton entanglement dynamics of the 
 Fenna-Matthews-Olson (FMO) pigment-protein  complex from the green sulfur bacteria
of the species, \textit{Prosthecochloris (P.) aestuarii} using typical
 values of the reservoir characteristics at cryogenic and physiological temperatures  \cite{lorenExpt}. 
Using  the time-convolutionless (TCL) projection operator 
technique and a Lorentzian spectral density of  phonon  reservoir,  
the evolution of the entanglement measure of  the  excitonic qubit $W$ states is 
evaluated,  showing  increased oscillations in
the entanglement  in the non-Markovian
regime.  The calculations are repeated for  
 the evolution dynamics of the Meyer-Wallach  measure of the
$N$-exciton multipartite state showing that  multipartite  entanglement can last up to at least 0.5 to 1.0 ps,
 between the Bchls of the FMO complex, in a decoherent environment. 

We have also considered quantum information processing  based on teleportation
and a complete protocol involving for the $GHZ$ and $W_{A}$ resource states, within
the context of photosynthetic systems.
The  teleportation fidelities associated with the $GHZ$ and $W_{A}$ resource states
associated with  the  excitonic qubits in the FMO complex of  \textit{P. aestuarii} show increased revivals in the fidelities  as the degree of non-Markovian
strength is increased. Unlike the $GHZ$ states, the $W$-like states appear to provide greater resilience
to environmental related decoherence during photosynthetic processes. 
Results indicate that quantum information processing involving teleportation followed by the
decodification tasks in $W_{A}$ states  of the FMO complex 
is likely to account for experimental results which show persistence of  coherent oscillations  
 at physiological temperatures. 

The results obtained thus far highlight  the importance role of several
 parameters ($\Delta \omega$ associated with the phonon 
reservoir correlation times, $\gamma_0$ associated with the exciton
relaxation times, $\delta$ associated with the detuning parameter)
 involved in  the  long coherence times  of
multipartite states in  the Fenna-Matthews-Olson  (FMO) pigment-protein complex of the  green sulfur bacteria. 
In particular the  non-Markovian regime appears best suited for occurrence of  large coherence times 
for the model system used in this study.
We note however that processes in   non-Markovian systems are not  well-understood, and further
analysis is need to confirm whether non-Markovianity, associated with  the backflow
of information from the reservoir  to the excitonic qubit system, is indeed 
instrumental in bringing about fast energy transport during photosynthesis.

Lastly, in large  light harvesting antennae systems, with intricate network connectivity, 
coherence between several  exciton states involving multipartite states can occur and 
contribute to the unique ability of these systems to attain robustness against decoherence.
The analysis of multipartite states 
carried out in this work can be extended to   estimate the multipartite entanglement
measures  in   photosynthetic systems with alternative forms of 
spectral functions associated with the environmental couplings . 
The  results obtained in this work may provide useful guidelines 
for future experimental work  involving the detection of multipartite entanglement
  in  photosynthetic systems. In particular,  advanced measurement tools
involving spectrally resolved, 4-wave mixing measurements \cite{segale}  may be utilized
to examine the remarkable resilience of  quantum correlations, and the role of multipartite states
 in   photosynthetic systems in contact with a decoherent environment.

\section{Acknowledgments}

This research was undertaken on the NCI National Facility in Canberra, Australia, which is
supported by the Australian Commonwealth Government.

\end{document}